\documentclass[12pt,fleqn]{article}

\usepackage{geometry} 
\usepackage{mathpazo} 

\usepackage[scaled]{helvet} 
\usepackage{courier} 
\normalfont
\usepackage[T1]{fontenc}

\usepackage{amsmath}
\usepackage{array}
\usepackage{bm}
\usepackage{graphicx}
\usepackage[numbers]{natbib}
\usepackage{path}
\usepackage{psfrag}
\usepackage{subfig}
\usepackage{chemarr, amssymb}
\usepackage{verbatim}
\usepackage{epstopdf}
\usepackage{float}

\usepackage{diagrams}
\usepackage{mathabx}
\usepackage{pdflscape}
\usepackage{comment}
\usepackage{wasysym}

\makeatletter
\newsavebox{\@brx}
\newcommand{\llangle}[1][]{\savebox{\@brx}{\(\m@th{#1\langle}\)}%
  \mathopen{\copy\@brx\kern-0.5\wd\@brx\usebox{\@brx}}}
\newcommand{\rrangle}[1][]{\savebox{\@brx}{\(\m@th{#1\rangle}\)}%
  \mathclose{\copy\@brx\kern-0.5\wd\@brx\usebox{\@brx}}}
\newcommand{\ffslash}[1][]{\savebox{\@brx}{\(\m@th{#1/}\)}%
  \mathclose{\copy\@brx\kern-0.5\wd\@brx\usebox{\@brx}}}
\makeatother

\usepackage{physics}

\newcommand{\lambdab}{{\boldsymbol \lambda}}  

\newcommand{\stirlingii}{\genfrac{\{}{\}}{0pt}{}}

\newtheorem{lemma}{Lemma}
\newtheorem{theorem}{Theorem}
\newtheorem{corollary}{Corollary}

\usepackage{outlines}

\usepackage[plainpages=false,pdfborder={0 0 0}]{hyperref}

\makeatletter
\let\@fnsymbol\@arabic
\makeatother

\begin{document}

\title{
Structural Commutation Relations for \\
Stochastic Labelled Graph Grammar \\
Rule Operators}

\author{Eric Mjolsness
\thanks{\,Department of Computer Science, University of California Irvine CA 92697.  Email: emj@uci.edu .} 
}

\maketitle

\begin{abstract}

We show how to calculate the operator algebra and the
operator Lie algebra of a stochastic labelled-graph grammar.
More specifically, we carry out a generic calculation 
of the product (and therefore the commutator)
of time-evolution operators for any two labelled-graph grammar rewrite rules,
where the operator corresponding to each rule is defined in terms
of elementary two-state creation/annihilation operators.
The resulting graph grammar algebra has the following properties:
(1) The product and commutator of two such operators 
is a sum of such operators with integer coefficients.
Thus, the algebra and the Lie algebra occurs entirely at the
structural (or graph-combinatorial) level of graph grammar rules,
lifted from the level of elementary creation/annihilation operators
(an improvement over 
[1], Propositions 1 and 2).
(2) The product of the off-diagonal (state-changing) parts of two such graph rule operators
is a sum of off-diagonal graph rule operators with non-negative integer coefficients.
(3) These results apply whether the semantics of a graph grammar rule
leaves behind hanging edges (Theorem 1), or removes hanging edges (Theorem 2).
(4) The algebra is constructive in terms of elementary two-state
 creation/annihilation operators (Corollaries 3 and 8).
 These results are useful because dynamical transformations of labelled graphs comprise
 a general modeling framework, and algebraic commutators of time-evolution operators
 have many analytic uses 
 including designing simulation algorithms and estimating their errors.

\end{abstract}

\section{Introduction}
\label{intro}

In ([1], 
Propositions 1 and 2) we 
showed
that the labelled-graph rewrite rule operator semantics specified 
there (in two versions, one without and one with hanging edge removal)
is contained within a somewhat larger operator algebra closed
under addition, scalar multiplication, and operator multiplication
(and hence under commutation, as in a Lie algebra).
The purpose of this paper is to show that the enlargement is not necessary.
So, under either semantics (hanging edges removed or not),
the vector space spanned by the 
graph rewrite rule operators previously defined form an
operator algebra and a Lie algebra among themselves.
In particular, the product of the state-changing portions of
two such operators can be written
as a sum of such operators with nonnegative integer weights,
and the full product and commutator of two such operators can be written
as a sum of such operators with integer weights.

These results occur within a larger scope discussed at length in [1],
including grammar-like or rule-based structured models of molecular complexes [2]
and of tissues with dividing cells [3, 4]
Potential applications include cytoskeletal dynamics in cellular and developmental biology,
neurobiology, and smart materials as well as the dynamics of
more abstract, non-spatial graphs in a wide variety of fields.

Given such state-changing operators $\hat{W}_r$ for the rules in a grammar,
the Master Equation for the stochastic dynamics is [5]
\begin{equation}
\begin{split}
{ d p \over d t} &= W \cdot p \quad \text{, where} \\
D_r &= \text{diag}({\mathbf 1} \cdot \hat{W}_r) \; ;  \quad 
& W_r = \hat{W}_r - D_r \; ; \quad &
W = \sum_r W_r \; ; \\
\end{split}
\label{MasterEqn}
\end{equation}
(generalizing [6, 7, 8] for stochastic chemical reaction networks),
and
where probability is defined over a suitable Fock space for varying numbers of graph nodes (with labels) and graph edges.
Here we assume this and related background as explained in [1, 9], 
for efficiency in calculating the main result (Section~\ref{advert_main_result}, \ref{sec_calculations}).

\section{Problem statement and main result}

\subsection{Graph grammar rule semantics}

In the following as described in [1,10,9], 
stochastic labelled graph grammar (SLGG)
rule  semantics with variables $X$ in the labels
is obtained by integrating
over a collection of rule variables $X$ that appear in graph labels $\lambda$;  as a special case,
 some $\lambda$ parameters can be constant as a function of $X$.
 Then

\begin{equation}
{\hat{W}}_{r} = \int d \mu_r(X)    {\hat{W}}_{r}({\lambdab}(X) , {\lambdab^{\prime }}(X)) 
\label{lambda_integral}
\end{equation}
where $\mu_r(X)$ is a suitable measure that could be discrete (so the integral becomes a sum)
or continuous.

Define
``$\sum_{\langle i_{1},\ldots i_{k}\rangle_{\neq} } \ldots $'' 
to be a sum over indices $ (i_{1},\ldots i_{k})$
constrained so that each $i_l$ is unequal to all the others,
in the simplest case (but see Section~\ref{hanging_edge_cleanup}) 
we define the time-evolution operator of a graph rewrite rule:
\begin{equation}
\begin{split}
{\hat{W}}_{r} \; =
{1 \over C_r(N_{\text{max free}})}
\int 
d X    \; \rho_{r}({\lambdab}(X) , {\lambdab^{\prime }}(X)) &
\sum \limits_{\langle i_{1},\ldots i_{k}\rangle_{\neq} }
 {\hat{a}}_{i_{1},\ldots i_{k}}(G^{r \; \text{out}})
 {{a}}_{i_{1},\ldots i_{k}}(G^{r \;  \text{in}})
 \end{split}
\label{eq1}
\end{equation}
where as explained in [1] the graph grammar rule operator
first annihilates all the edges and labelled nodes in the incoming graph
and then, 
but uninterruptibly and with zero delay, 
creates the  corresponding elements of the outgoing graph:
\begin{equation}
\begin{split}
{\hat{a}}_{i_{1},\ldots i_{k}}(G^{\prime}) &=
{\hat{a}}_{i_{1},\ldots i_{k}}(G^{\prime}_{\text{links}})
{\hat{a}}_{i_{1},\ldots i_{k}}(G^{\prime}_{\text{nodes}}) 
\\ &=
\Bigg [ \prod \limits_{s^{\prime}, t^{\prime} \in \operatorname{rhs}( r) } {\left( {\hat{a}}_{i_{s^{\prime}}
i_{t^{\prime}}}\right) }^{{g^{\prime }}_{s^{\prime} \; t^{\prime} }} \Bigg] 
\Bigg[  \prod \limits_{v^{\prime} \in
\operatorname{rhs}(r) }  
{\hat{a}}_{i_{v^{\prime}} {\lambda^{\prime }}_{v^{\prime}}}\Bigg] 
\\ &=
\Bigg [ \prod \limits_{(s^{\prime}, t^{\prime}) \in G^{\prime}_{\text{links}} } 
{\hat{a}}_{i_{s^{\prime}} i_{t^{\prime}}}
\Bigg] 
\Bigg[  \prod \limits_{v^{\prime} \in
G^{\prime}_{\text{nodes}} }  
{\hat{a}}_{i_{v^{\prime}} {\lambda^{\prime }}_{v^{\prime}}}\Bigg] 
 \\ 
  {{a}}_{i_{1},\ldots i_{k}}(G) & = 
{{a}}_{i_{1},\ldots i_{k}}(G_{\text{links}})
{{a}}_{i_{1},\ldots i_{k}}(G_{\text{nodes}}) 
\\ &=
\Bigg[ \prod \limits_{s, t \in \operatorname{lhs}( r) } {\left( a_{i_{s}
i_{t}}\right) }^{g_{s \; t}} \Bigg] 
\Bigg[  \prod \limits_{v \in \operatorname{lhs}(r) }  
a_{i_{v} \lambda_{v}} \Bigg] . %
\\ &=
\Bigg[ \prod \limits_{(s, t) \in G_{\text{links}} } 
a_{i_{s}
i_{t}}
\Bigg] 
\Bigg[  \prod \limits_{v \in G_{\text{nodes}} }  
a_{i_{v} \lambda_{v}} \Bigg] . %
\end{split}
\label{eq2}
\end{equation}
The sets $\text{lhs}_r$ and $\text{rhs}_r$ comprise the nodes or vertices
in the left hand side and ride hand side graphs, $G$ and $G^{\prime}$, 
with adjacency matrices $g$ and $g^{\prime}$, of rule $r$.
The factor of ${1 / C_r(N_{\text{max free}})}$ in Equation~(\ref{eq1})
will be discussed in Section~\ref{normalization}.

In order to specify labelled graphs by ordinary syntactic means,
we have imposed a {\it numbering} on the nodes in
$\text{lhs}_r$ and $\text{rhs}_r$ as discussed further in [1];
this numbering allows us to use adjacency matrices
$g$ and $g_{\prime} $ to define the graph links or edges,
to map labels to nodes, and to define disjoint unions of nodes
where needed.

Thus
from Equations~(\ref{eq1}) and (\ref{eq2}),
\begin{equation}
\begin{split}
{\hat{W}}_{r} \; =
{1 \over C_r(N_{\text{max free}})}
\int 
d \mu_r(X)    \; \rho_{r}({\lambdab}(X) , {\lambdab^{\prime }}(X)) &
\sum \limits_{\langle i_{1},\ldots i_{k}\rangle_{\neq} }
\Bigg [ \prod \limits_{s^{\prime}, t^{\prime} \in \operatorname{rhs}( r) } {\left( {\hat{a}}_{i_{s^{\prime}}
i_{t^{\prime}}}\right) }^{{g^{\prime }}_{s^{\prime} \; t^{\prime} }} \Bigg] 
\Bigg[  \prod \limits_{v^{\prime} \in
\operatorname{rhs}( r) }  
{\hat{a}}_{i_{v^{\prime}} {\lambda^{\prime }}_{v^{\prime}}}\Bigg] 
 \\ & \times
\Bigg[ \prod \limits_{s, t \in \operatorname{lhs}( r) } {\left( a_{i_{s}
i_{t}}\right) }^{g_{s \; t}} \Bigg] 
\Bigg[  \prod \limits_{v \in \operatorname{lhs}(r) }  
a_{i_{v} \lambda_{v}}  \Bigg] . %
\end{split}
\label{eq3}
\end{equation}
Also we have the product
\begin{equation}
\begin{split}
{\hat{W}}_{r_2} {\hat{W}}_{r_1} \; =& \; 
{1 \over C_{r_1}(N_{\text{max free}})} {1 \over C_{r_2}(N_{\text{max free}})}
\int \int
d \mu_{r_1}(X_1)    d \mu_{r_2} (X_2) 
  \; \rho_{r_1}({\lambdab_1}(X_1) , {\lambdab_1^{\prime }}(X_1)) 
    \; \rho_{r_2}({\lambdab_2}(X_2) , {\lambdab_2^{\prime }}(X_2)) &
\\ & %
\sum \limits_{\langle j_{1},\ldots j_{k_2}\rangle_{\neq} }
\sum \limits_{\langle i_{1},\ldots i_{k_1}\rangle_{\neq} }
 {\hat{a}}_{j_{1},\ldots j_{k_2}}(G^{r_2 \; \text{out}}_{\text{links}})
{\hat{a}}_{j_{1},\ldots j_{k_2}}(G^{r_2 \; \text{out}}_{\text{nodes}}) 
{{a}}_{j_{1},\ldots j _{k_2}}(G^{r_2 \; \text{in}}_{\text{links}})
{{a}}_{j_{1},\ldots j_{k_2}}(G^{r_2 \; \text{in}}_{\text{nodes}}) 
 \\  &  \times
 {\hat{a}}_{i_{1},\ldots i_{k_1}}(G^{r_1 \; \text{out}}_{\text{links}})
{\hat{a}}_{i_{1},\ldots i_{k_1}}(G^{r_1 \; \text{out}}_{\text{nodes}}) 
 {{a}}_{i_{1},\ldots i_{k_1}}(G^{r_1 \; \text{in}}_{\text{links}})
{{a}}_{i_{1},\ldots i_{k_1}}(G^{r_1 \; \text{in}}_{\text{nodes}}) \; ,
\end{split}
\label{eq4}
\end{equation}
and consequently
\begin{equation}
\begin{split}
{\hat{W}}_{r_2} {\hat{W}}_{r_1} \; =& \; 
{1 \over C_{r_1}(N_{\text{max free}})} {1 \over C_{r_2}(N_{\text{max free}})}
\int \int
d \mu_{r_1}(X_1)    d \mu_{r_2} (X_2) 
  \; \rho_{r_1}({\lambdab_1}(X_1) , {\lambdab_1^{\prime }}(X_1)) 
    \; \rho_{r_2}({\lambdab_2}(X_2) , {\lambdab_2^{\prime }}(X_2)) &
\\ & %
\sum \limits_{\langle j_{1},\ldots j_{k_2}\rangle_{\neq} }
\sum \limits_{\langle i_{1},\ldots i_{k_1}\rangle_{\neq} }
 {\hat{a}}_{j_{1},\ldots j_{k_2}}(G^{r_2 \; \text{out}}_{\text{links}})
 \Big[
{{a}}_{j_{1},\ldots j _{k_2}}(G^{r_2 \; \text{in}}_{\text{links}})
 {\hat{a}}_{i_{1},\ldots i_{k_1}}(G^{r_1 \; \text{out}}_{\text{links}})
 \Big]
 {{a}}_{i_{1},\ldots i_{k_1}}(G^{r_1 \; \text{in}}_{\text{links}})
 \\  &  \times
{\hat{a}}_{j_{1},\ldots j_{k_2}}(G^{r_2 \; \text{out}}_{\text{nodes}}) 
 \Big[
{{a}}_{j_{1},\ldots j_{k_2}}(G^{r_2 \; \text{in}}_{\text{nodes}}) 
{\hat{a}}_{i_{1},\ldots i_{k_1}}(G^{r_1 \; \text{out}}_{\text{nodes}}) 
 \Big]
{{a}}_{i_{1},\ldots i_{k_1}}(G^{r_1 \; \text{in}}_{\text{nodes}})  \; .
\end{split}
\label{eq4_1}
\end{equation}

\subsection{Problem statement}

Up to equivalence, how can the product of two graph grammar rewrite rule operators be expressed
in terms of a sum of such operators? 

Likewise for the commutator of such operators?

\subsection{Equivalence of rule operators}

Two models defined by the Master Equation (ME) will be
``equivalent'' if their state variables can be identified so
that solutions of the Master Equation
are identical in all statistically observable respects:
in all moments of all number operators at all choices of observation
times. If $\alpha$ indexes the observable 
numbers  $n_\alpha$ of objects and $N_\alpha$ is the corresponding
number operator, then we can read out a broad range of
joint probabilities with the moments of Kronecker delta functions:
\begin{equation}
\text{Pr}_{\text{\tiny ME}}([n_{\alpha(q)}(t_q)|q]) = \Big \langle \prod_q \delta (N_{\alpha(q)}(t_q)-n_{\alpha(q)} I_{\alpha(q)}) \Big \rangle_{\text{\tiny ME}}
\label{particle-equivalence}
\end{equation}
As the operative definition of {\it equivalence}, we demand equality of all such moments.
Other observables 
$\langle f([N_{\alpha(q)}(t_q) | q ] ) \rangle_{\text{\tiny ME}} $ 
(where $f$ is applied componentwise to diagonal matrices)
follow from Equation~(\ref{particle-equivalence}) as a linear basis.

\subsection{Main result}
\label{advert_main_result}

After a calculation and several arguments, the main result will take the form
of an operator algebra equivalence that turns products of graph rewrite operators
into sums of other graph rewrite operators:

\begin{equation}
\boxed{
{\hat{W}}_{G^{r_2 \; \text{in}}  {\rightarrow} G^{r_2 \; \text{out}}  }
    {\hat{W}}_{G^{r_1 \; \text{in}}  {\rightarrow} G^{r_1 \; \text{out}}  }
	 \; \simeq \;
   \sum_{
   	\vbox{ 
		\hbox{\scriptsize $H \subseteq G^{r_1 \, \text{out} } \simeq \tilde{H} \subseteq G^{r_2 \, \text{in} }$ }
		 \hbox{\scriptsize \quad \;\;  | edge-maximal}
	}
     }
	\sum_{h: H \overset{\text{\tiny 1--1}}{  \hookrightarrow } \tilde{H}}
    \quad
{\hat{W}}_{G^{1;2 \; \text{in}}(\tilde{H}) 
	\underset{h}{\rightarrow}
	G^{1;2 \; \text{out}} (H) } 
}
\label{main_result_structural}
\end{equation}
where the new labelled graphs, roughly given by
\begin{equation}
\begin{split}
G^{1;2 \; \text{in}}(\tilde{H}) &= G^{r_1 \; \text{in}} \cup (G^{r_2 \; \text{in}}  \setminus \tilde{H}) \\
G^{1;2 \; \text{out}}(H) &= G^{r_2 \; \text{out}}  \cup (G^{r_1 \; \text{out}} \setminus H) \; ,
\end{split}
\end{equation}
and their labelled-graph overlap
will be defined more carefully
in Section~\ref{sec_calculations}.

This result will be shown both without (Theorem 1, Section~\ref{calculation-no-cleanup})
and with (Theorem 2, Section~\ref{calculation-cleanup})
hanging edge cleanup semantics.
First we discuss some operator algebra calculational techniques and strategies we use,
without claiming any optimality for them.

In addition, in the course of proving these two theorems we exhibit in each case a 
{\it constructive mapping}  (Corollaries 3 and 8)
from the graph rewrite rule operator algebra semantics
to the elementary bitwise (two-state) operator algebras of Section~\ref{bitwise_op_alg}.

\section{Techniques}

\subsection{Normalization}
\label{normalization}

The factor of ${1 / C_r(N_{\text{max free}})}$ in Equation~(\ref{eq1})
accounts for a large number of equivalent states that could result from
a rule firing, whose weight should add up to ${\cal O}(1)$. It
reflects the fact that
in operator algebra formalism reaction rates naturally follow the law of mass action,
so that if (as one would hope) a 
large number $N_{\text{max free}}$
of unallocated 
node indices are available for creating new graph content
then the net rate of creation for that content is proportionately very high;
yet this factor should instead be unimportant, so we scale it out.
Roughly, $C_r(N_{\text{max free}})$ should be $N_{\text{max free}} ! / ((N_{\text{max free}})- m_r)!$
where $m_r$ is the number of new nodes $|G^{r \; \text{out}}_{\text{nodes}} \setminus G^{r \; \text{in}}_{\text{nodes}}|$
appearing in the output graph but not the input graph.
However, $N_{\text{max free}}$ should 
be much larger than $m_r$ so that it does
not change appreciably when graph nodes are created or destroyed,
in which case  $C_r(N_{\text{max free}}) \simeq (N_{\text{max free}})^{m_r}$
with equality in the $N_{\text{max free}} \rightarrow +\infty$ limit.
Then in the limit $C_r$ is ``multiplicative'' for additive $m_r$
(i.e. $(N_{\text{max free}})^{m_{r_1}} (N_{\text{max free}})^{m_{r_2}} = (N_{\text{max free}})^{m_{r_1}+m_{r_2}} $ 
as we assume for Theorems 1 and 2 below).
Alternatively, $C_r$ could be held constant by an index allocation mechanism
such as that described in Section~\ref{CMA}.
(Thus, one could invent a memory gatekeeping mechanism
similar to "malloc" in C, "new" in C++, and "gensym" in Lisp, but expressed in operator algebraic notation
for allocating one block of indices at a time,
at the risk of some degree of unnecessary serialization.)
A useful limit of this route is to set $N_{\text{max free}}=1, C_r=1$ 
(also multiplicative) by imposing
a unique choice of new, unique index value for each new node generated in
each rule firing; this method requires a suitable choice function.

\subsection{Operator algebra techniques}

The expressions $[\ldots]$ in square brackets 
in Equation~(\ref{eq4_1})
need to be restored to normal order, with annihilators 
$a_{\alpha} $ to the right of (preceding) creation operators $\hat{a}_{\alpha}$ .

\subsubsection{Elementary operators' algebra}
\label{bitwise_op_alg}

To do this systematically we need various operator rules for $2x2$ elementary operators:
\begin{subequations}
\begin{align}
\hat{a} &=\left( \begin{array}{cc}
 0 & 0 \\
 1 & 0
\end{array}\right)  , a=\left( \begin{array}{cc}
 0 & 1 \\
 0 & 0
\end{array}\right)
 \; \text{implies} \;
\\
\hat{a}a &=N\equiv \left( \begin{array}{cc}
 0  & 0  \\
 0 & 1
\end{array}\right) \; , \quad 
a \hat{a} = Z \equiv I - N 
=
\left( \begin{array}{cc}
1  & 0  \\
 0 & 0
\end{array}\right)\; , \; \text{and} \\ 
[a_\alpha,\hat{a}_\beta] &= \delta_{\alpha \beta}(I_\alpha- 2 N_\alpha ) I
\quad \quad {\text{Alternative for normal form calcs:}} \\
a_\alpha \hat{a}_\beta &= \hat{a}_\beta a_\alpha  - 2 \delta_{\alpha \beta} \hat{a}_\alpha a_\alpha + \delta_{\alpha \beta} I_{\alpha} \\
a_\alpha \hat{a}_\beta & = (1- \delta_{\alpha \beta}) \hat{a}_\beta a_\alpha  + \delta_{\alpha \beta} Z_{\alpha}
%
\end{align}
\label{binary_algebra_1}
\end{subequations}
Then
\begin{equation}
\begin{split}
\hat{a}_{\alpha}^2  =& 0 = a_{\alpha}^2 \\
N_{\alpha} \equiv & \hat{a}_{\alpha} a_{\alpha} \quad \quad \text{\small (diagonal)} \\
Z_{\alpha} \equiv & I_{\alpha} - N_{\alpha}  \quad \text{\small (diagonal)} \\
a_{\alpha} \hat{a}_{\beta} =& (1-\delta_{\alpha \beta}) \hat{a}_{\beta} a_{\alpha} + \delta_{\alpha \beta} Z_{\alpha}  \\
\end{split}
\quad \text{and} \quad
\begin{split}
N_{\alpha} N_{\alpha}  =&  N_{\alpha}  \\
Z_{\alpha}  Z_{\alpha} =&  Z_{\alpha}  \\
\end{split}
\quad \text{and} \quad
\begin{split}
Z_{\alpha} a_{\alpha} =&  a_{\alpha} \\
 a_{\alpha} Z_{\alpha} =&  0 \\
Z_{\alpha} \hat{a}_{\alpha} =&  0\\
\hat{a}_{\alpha} Z_{\alpha} =&  \hat{a}_{\alpha} \\
\end{split}
\label{eq5}
\end{equation}

An extra multiplicative algebra sector governs the erasure operator $E \equiv Z+a$:
\begin{equation}
\begin{split}
E_{\alpha} & \equiv \Pi_{0 \, \alpha} \equiv Z_{\alpha} + a_{\alpha}  \\
E_{\alpha} a_{\alpha} &=  a_{\alpha} \\
 a_{\alpha} E_{\alpha} &=  0 \\
E_{\alpha} \hat{a}_{\alpha} &=  Z_{\alpha} \\
\hat{a}_{\alpha} E_{\alpha} &=  \Pi_{1 \, \alpha} \\
\end{split}
\quad \text{and} \quad 
\begin{split}
\Pi_{1 \, \alpha} &\equiv   \hat{a}_{\alpha} + N_{\alpha}  \\
\Pi_{1 \, \alpha} a_{\alpha} &=  N_{\alpha} \\
 a_{\alpha} \Pi_{1 \, \alpha} &=  \Pi_{0 \alpha} \\
\Pi_{1 \, \alpha} \hat{a}_{\alpha} &=  \hat{a}_{\alpha} \\
\hat{a}_{\alpha} \Pi_{1 \, \alpha} &=  0 \\
\end{split}
\label{eq8}
\end{equation}

In order to control the signs of integer-valued
weights in  operator products,
we observe the following:
For creation/annihilation operators pertaining to graph edges,
including those making up the edge erasure operators $E_{i_p \; i}$ and $E_{i \; i_q}$,
using e.g. Equation~(\ref{binary_algebra_1}e) rather than (\ref{binary_algebra_1}d)
removes the explicit negative signs from the algebra
by introducing matrix $Z_{i_p \; i_q}$ which has nonnegative entries.

This algebra governs the graph edge creation and annihilation operators,
for which $\alpha = (i,j)$.
It does not apply directly to the node label creation and annihilation operators,
except as targets of an operator homomorphism to be described next.
For this homomorphism the elementary bitwise operators obeying the algebra
above will be denoted ``$b$'' rather than ``$a$''.

\subsection{Operator Algebra homomorphisms}

A homomorphism of operator algebras is
defined here as
a mapping from one operator algebra to another
that preserves the basic algebraic operations:
finite sums, scalar multiplication, and finite products of operators.
It is thus a ring homomorphism, for a ring of linear operators that act on a vector space.
In our case the vector space is a Fock space 
capable of hosting classical probability distributions
[5, 9, 1].
If the operator algebra homomorphism is also injective, it could be called an ``embedding''.

\subsubsection{Winner Take All (WTA or 1-Hot) Encoding of Labels}

We can enforce a winner-might-take-all logic of labels either by fiat using axioms:
\begin{equation}
\begin{split}
a_{i , \, \lambda} a_{i , \,\lambda^{\prime}} =& 0 \\
\hat{a}_{i ,\,\lambda} \hat{a}_{i , \,\lambda^{\prime}} =& 0 \\
{a}_{i ,\,\lambda} \hat{a}_{i , \,\lambda^{\prime}} =&  \delta_{\lambda \,\lambda^{\prime}} Y_{i , \,\lambda^{\prime} } .
\end{split}
\label{eq7}
\end{equation}
where $N^{(a)}_{i , \,\lambda^{\prime}}$ and $Y_{i , \,\lambda^{\prime}}$ are 
diagonal in the number basis and idempotent, satisfying
\begin{equation}
\begin{split}
Y_{\alpha} a_{\alpha} =&  a_{\alpha} \\
 a_{\alpha}  Y_{\alpha} =&  0 \quad   \\
Y_{\alpha} \hat{a}_{\alpha} =&  0\\
\hat{a}_{\alpha} Y_{\alpha} =&  \hat{a}_{\alpha} \\
\end{split}
\quad \text{and} \quad
\begin{split}
Y_{\alpha} a_{\beta} =&  a_{\alpha} Y_{\beta} \quad \text{ for } (\alpha \neq \beta) \\
Y_{\alpha} \hat{a}_{\beta} =&  \hat{a}_{\beta} Y_{\alpha} \quad \text{ for }  (\alpha \neq \beta)\\
Y_{\alpha} Y_{\alpha} =&  Y_{\alpha} \\
\end{split}
\label{Y_algebra}
\end{equation}
for $\alpha = (i,\lambda)$ as appropriate for node labels.
Likewise for $N$:
\begin{equation}
\begin{split}
N_{\alpha} a_{\alpha} =&  0  \\
 a_{\alpha}  N_{\alpha} =&  a_{\alpha} \quad \\
N_{\alpha} \hat{a}_{\alpha} =&   \hat{a}_{\alpha} \\
\hat{a}_{\alpha} N_{\alpha} =& 0 \\
\end{split}
\quad \text{and} \quad
\begin{split}
N_{\alpha} a_{\beta} =&  a_{\alpha} N_{\beta} \quad \text{ for } (\alpha \neq \beta) \\
N_{\alpha} \hat{a}_{\beta} =&  \hat{a}_{\beta} N_{\alpha} \quad \text{ for }  (\alpha \neq \beta)\\
N_{\alpha} N_{\alpha} =&  N_{\alpha} \\
\end{split}
\label{N_algebra}
\end{equation}
and $N_\alpha Y_\alpha = 0 = Y_\alpha N_\alpha$; also $N_\alpha Y_\beta = Y_\beta N_\alpha$.

Alternatively, we can ground this WTA algebra in terms of 
the usual elementary 0/1-valued states using
the 0/1-winner mapping
\begin{equation}
\begin{split}
a_{i , \, \lambda}  &= {\hat b}_{i , \,\varnothing}  b_{i , \,\lambda}   \\
\hat{a}_{i , \, \lambda}  &=   \hat{b}_{i , \,\lambda} b_{i , \, \varnothing}  \quad 
\end{split}
\label{eq7_3}
\end{equation}
in which the $b,\hat{b}$ operators obey the bitwise algebra above,
and they {\it also} by induction obey the WTA/one-hot subspace constraint
imposed by initial condition and preserved by operators constructed from $a,\hat{a}$:
\begin{equation}
\begin{split}
N_{i , \, \varnothing}  + \sum_\lambda N^{(b)}_{i , \,\lambda} &\simeq I \\
b_{i , \, \varnothing}  b_{i , \,\lambda} &\simeq   0  \simeq b_{i , \,\lambda} b_{i , \, \varnothing} \quad , \\
b_{i , \,\lambda} b_{i , \,\lambda^{\prime}} &\simeq   0  \quad .
\end{split}
\label{eq7_3_2}
\end{equation}
In the number basis for $b$, these equivalences follow from
the initialization and inductive preservation of
\begin{equation}
\begin{split}
n_{i , \, \varnothing}  + \sum_\lambda n^{(b)}_{i , \,\lambda} &=1  \\
n_{i , \, \varnothing} , \, n^{(b)}_{i , \,\lambda} &\in \{0,1\}  \\
\end{split}
\end{equation}
so that
 $n_{i , \, \varnothing}  n^{(b)}_{i , \,\lambda}=0$
 and $\lambda \neq \lambda^\prime \implies n^{(b)}_{i , \,\lambda}  n^{(b)}_{i , \,\lambda^\prime}=0$;
 then use $b_\alpha | \ldots n_\alpha \ldots\rangle = n^{(b)}_\alpha | \ldots (n_\alpha-1) \ldots \rangle$ .

Using this algebra for $b,\hat{b}$ and the operator algebra homomorphism to $a,\hat{a}$
induced by Equation~(\ref{eq7_3}), then the $a,\hat{a}$ algebra of
Equations~(\ref{eq7}),(\ref{Y_algebra}), and (\ref{N_algebra}) 
(interpreting $N$ in (\ref{eq7})-(\ref{N_algebra}) as $N^{(a)}$ below, not as $N^{(b)}$)
can be verified by direct computation. 
We find the additional homomorphism mappings 
to the bitwise ``$b$'' algebra
for $Y$ and $N^{(a)}$:
\begin{equation}
\begin{split}
N^{(a)}_{i, \lambda} &= N^{(b)}_{i, \lambda} Z_{i, \varnothing} 
    =   \hat{b}_{i , \,\lambda} b_{i , \,\lambda}   b_{i , \, \varnothing}  {\hat b}_{i , \,\varnothing}  \\
Y_{i, \lambda} &= Z^{(b)}_{i, \lambda} N_{i, \varnothing} 
    =   b_{i , \,\lambda}  \hat{b}_{i , \,\lambda} {\hat b}_{i , \,\varnothing}   b_{i , \, \varnothing}   \quad 
\end{split}
\label{more_homs}
\end{equation}

Of course operators indexed by nodes $i \neq j$ all commute.
Combined with the last line of Equation~(\ref{eq7}), this fact produces
a major calculational tool for nodes in the form of the following key commutation relation:
\begin{equation}
\boxed{
a_{j \, , \, \lambda} \hat{a}_{i \, , \, \lambda^{\prime}} 
= (1-\delta_{i j} ) \hat{a}_{i \, , \, \lambda^{\prime}} a_{j \, , \, \lambda} +\delta_{i j}  \delta_{\lambda \lambda^{\prime}}Y_{j \, , \, \lambda^{\prime}}
}
\label{key_comm}
\end{equation}
This relation differs from Equation~(\ref{binary_algebra_1}e) 
in producing fewer nonzero results, 
so it is more constraining,
and a slightly different diagonal operator $Y$ obeying the same algebra for node labels
(in Equation~(\ref{Y_algebra})) as $Z$ (in Equation~(\ref{eq5})) does for edges.
Equation~(\ref{binary_algebra_1}e)  however still governs edge operators.

Thus we reach sufficient multiplicative information on $\{a, \hat{a}, N, Z, Y, E, \Pi_1 \}$ 
in principle  to compute all products of $\hat{W}$ operators.

\subsubsection{Controlled index allocation}
\label{CMA}

Each graph rewrite rule may introduce new graph nodes not already present.
The graph rewrite rule algebra will be simpler if these can be modeled with
fresh node indices  $i$ not previously used - even if some 
further algebra homomorphism and
remapping not undertaken
here actually reuses old, deallocated graph node indices.
(Edge index pairs will necessarily be fresh - heretofore unused 
- if at least one of their node indices is fresh.)
Here we just seek to express algebraically a continual, parallelism-compatible 
supply of fresh indices. Choose an index block size $B$ that is large enough to encompass
the new nodes of any rule we consider. 
The chosen $B$ could even be countably infinite, e.g. if we use a diagonal raster
traversal of $B$ and the infinite collection of blocks needed;
however in Theorem 2 we will assume $B$ is finite.
Index the blocks needed by
$\mu \in M$ where $M$ is a countably infinite tree
of finite maximum branching degree, and let $\phi \subseteq M$ denote 
a frontier in $M$:
the collection of next blocks available for allocation,
whose ancestors have all been allocated.

As a special case,
if $M$ is a graph isomorphic to the integers with
succession (${\mathbb N}^+$)
as the tree relationship, then this scheme will force serial computation;
but an average branching degree even slightly greater than 1 permits parallelism.

Each block $\tau$ has binary variables  $A_\tau \in \{0,1\}$
taking the value 1 if and only if block $\tau$ is ''allocated'' or ``alive''
(in which case all of $\tau$'s ancestors mush also be alive), and
$F_\tau \in \{0,1\}$ taking the value 1 if and only if block $\tau$ is in the current
frontier $\phi$, in which case $A_\tau=1$ but all of $\tau$'s children
must be unallocated ($A_{\sigma \in \text{children}(\tau)}=0$). 
These binary variables $F$ have creation/annihilation operators
$\hat{b}^{\text{ind}}_\tau$ and $b^{\text{ind}}_\tau$.
Then $| \phi | = \sum_{\tau \in M} F_\tau$. 

We will assume that nodes $i$ which have never been allocated in a memory block
all obey the initial condition
that $n_{i \, , \varnothing}=1$ and $n_{i \, , \lambda}=0$ for other labels $\lambda$
and inductively have no way of changing until the memory block $\tau$ containing $i$ is allocated;
and likewise all the edge numbers $n_{i j}$ and $n_{ j i}$ involving node $i$ are all initialized to zero
and inductively have no way of changing until the memory blocks $\sigma,\tau$ containing
 $i$ and $j$ respectively are both allocated.

Let $\text{Ch}(\tau)$ be the set of child blocks of memory block $\tau$ in $M$.
Then the combined operator 
\begin{equation}
\text{Advance}_\tau \equiv
\Big[ \prod_{\sigma \in \text{Ch}(\tau)}  
\hat{b}^{\text{ind}}_\sigma \Big]
b^{\text{ind}}_\tau
\label{advance_op}
\end{equation}
could be used to advance the frontier $\phi$ of allocated memory under a single rule firing.
($M$ could even be permitted to be a directed acyclic graph, if the child operator $\hat{b}^{\text{ind}}_\sigma$
in the product in Equation~(\ref{advance_op}) 
is replaced by $(\hat{b}^{\text{ind}}_\sigma+N^{\text{ind}}_\sigma)$.
Then child memory blocks that are already alive and in the frontier are permitted, and remain that way.)
If we initialize the aliveness and frontier at the root of the tree and maintain it by
Equation~(\ref{advance_op})  inductively, then we can take
\begin{equation}
\begin{split}
a^{\text{ind}}_\tau &= b^{\text{ind}}_\tau \\
\hat{a}^{\text{ind}}_\tau
& =
\hat{b}^{\text{ind}}_\tau
\end{split}
\label{memvblschange_a}
\end{equation}
More conservatively we could continually check that old memory is not about to be
reused incorrectly:
\begin{equation}
\begin{split}
a^{\text{ind}}_\tau &= b^{\text{ind}}_\tau \\
\hat{a}^{\text{ind}}_\tau
& =
\hat{b}^{\text{ind}}_\tau
\Big( \prod_{\sigma \in \text{ancestors}(\tau)} 
Z^{\text{ind}}_\sigma \Big)
\end{split}
\label{memvblschange_b}
\end{equation}

The index allocation frontier maps to parallel computational architectures in which time can be local,
for example time can be a spacelike foliation of spacetime that respects signal propagation delays.

Now the idea is that rule-firing operators $\hat{W}_r$ will act also in the index allocation space,
using and then
advancing the frontier $\phi$ of blocks $\tau$ from which newly allocated graph nodes $i$ can be drawn.
Denoting by $\hat{W}_{r, \tau}$ the variant of $\hat{W}_r$ that draws all newly allocated nodes $i$
from block $\tau$ (the block size $B$ being always large enough for this),
then 
\begin{equation}
\begin{split}
{\hat{W}}_{r} \; \equiv & \; 
\sum_{\tau \in M } 
\Big[ \prod_{\sigma \in \text{Ch}(\tau)}  
\hat{a}^{\text{ind}}_\sigma \Big]
{\hat{W}}_{r\tau}
a^{\text{ind}}_\tau
{{1}\over{| \phi  |}} .
\end{split}
\label{block_operator}
\end{equation}
where 
all such expressions as $\tau$ varies are regarded as equivalent
owing to index permutation invariance and operator linearity.
In the special case $M={\mathbb N}^+$, $\phi=\{ \tau \}, |\phi| =1$ and this operator becomes
\begin{equation}
\begin{split}
{\hat{W}}_{r} \; \equiv & \; 
\sum_{\tau \in M } 
\hat{a}^{\text{ind}}_{\tau + 1} 
{\hat{W}}_{r\tau}
a^{\text{ind}}_\tau
\end{split}
\label{block_operator_serial}
\end{equation}
(cf. [11], 
a quantum version that adds in the time-reversal Hermitian conjugate of all transitions)
which is the form we will assume.

 With more complex dynamics one could try try to ensure that 
 in Equation~(\ref{block_operator})
 the $|\phi|$, size of the frontier,
 is constant or nearly constant in time, and move its inverse to the left of the $\sum_\tau$ above. 
For example $M$ could be a root node connected to the zero nodes of $|\phi|$ half-infinite chains 
each isomorphic to the integers under succession.
 Alternatively one could track the relationship
 between simulated and computational time.
In what follows we'll assume one of these options has been taken,
so that the factor of $1/|\phi|$ is the same for all rules, 
treat the general $M$ case as {\it equivalent} (using $\simeq$ as previously defined)
to the special case $M={\mathbb N}^+$, $\phi=\{ \tau \}, |\phi| =1$ 
that we assume in the calculations that follow.

Similar ``aliveness'' variables in quantitative grammar models have
been used in [3] and [12],  
along with winner-take-all variable subset constraints,
though without the operator algebra framework.
Controlled index allocation could be related in a computational implementation to controlled memory allocation.

\subsubsection{Hanging edge cleanup}
\label{hanging_edge_cleanup}

Another elaboration of rule operators $\hat{W}_r$ can clean up hanging edges 
that may otherwise be left behind by a rule firing:

\begin{equation}
\begin{split}
\hat{W}_r^{\text{cleaned}} 
&= 
  \Big( 
  	\prod_{k_1 \in L_r \setminus R_r}
 	 \prod_{k_2 \in {\cal U} } 
  E_{k_1 k_2} E_{k_2 k_1}
  \Big)
  \hat{W}_r^{\text{bare}} \\
& \simeq
  \Big( 
  	\prod_{(k_1,k_2)  \in {\cal S} }
  	E_{k_1 k_2} 
  \Big)
  \Big( 
  	\prod_{(k_1,k_2)  \in {\cal S} }
	E_{k_2 k_1}
  \Big)
  \hat{W}_r^{\text{bare}} \\
\end{split}
\label{Edge_cleanup}
\end{equation}
where ${\cal S}$ is the set of indices specified by
\begin{equation}
  	{\cal S} = [(L_r \setminus R_r) \; \times \; 
		 {\cal U}_{A^*} ] 
\label{Edge cleanup}
\end{equation}
where ${\cal U}_{A*}$ = all node indices that have ever been allocated in a memory block,
hence all memory-live node indices,
and ${\cal U}$ = the whole universe of node indices, so that $ {\cal U}_{A^*} \subseteq {\cal U}$.
The second line
in Equation~(\ref{Edge_cleanup})
 is {\it equivalent}, $\simeq$ to the first because
as discussed above,
unallocated $k_2$ indices
inductively have $n_{k_1 k_2}=0=n_{k_2 k_1}$, 
and the erasure operator does nothing
(is equivalent to the identity operator) in that case.
The reason for including this restriction in the definition of ${\cal S}$ is that, 
for rules with finite graphs and index allocation with finite block size
and after any finite number of rule firings, ${\cal U}_{A^*} $ is finite
and both factors of the set ${\cal S}$ are finite, so ${\cal S}$ itself is finite;
only a finite amount of cleanup work needs to be done for each rule firing.
We will use this assumption in the proof of Theorem 2.

In the next section we will use the notation
${\cal P}_\chi = [ ( L_\chi \setminus R_\chi ) \times {\cal U}] $
for the predicate that designates the 
possibly infinite superset of
index set ${\cal S}$ above, that pertains to the top line of Equation~(\ref{Edge_cleanup}).

In greater detail the hanging-edge removal semantics 
as specified less formally in the top line of Equation~(\ref{Edge_cleanup})
is  
\begin{equation}
\begin{split}
{\hat{W}}_{r} \; =
{1 \over C_r(N_{\text{max free}})}
\; \rho _{r}({\lambdab} , {\lambdab^{\prime }}) &
\sum \limits_{\langle i_{1},\ldots i_{k}\rangle_{\neq} }
E_{\text {cleanup}}(G^r)
 {\hat{a}}_{i_{1},\ldots i_{k}}(G^{r \; \text{out}})
 {{a}}_{i_{1},\ldots i_{k}}(G^{r \;  \text{in}})
 \end{split}
\label{Edge_cleanup_full}
\end{equation}
where
\begin{equation}
\begin{split}
E_{\text {cleanup}}(G^{r \; \text{in}},G^{r \; \text{out}})
   &=
\Bigg[ 
 \Big(  \prod \limits_{p \in G^{r \; \text{in}}_{\text{nodes}} \setminus G^{r \; \text{out}}_{\text{nodes}}  } 
 	\; \prod_{ i  \in {\cal U} 
		} E_{i_{p} \; i}  \Big) 
  \Big(  \prod \limits_{p \in  G^{r \; \text{in}}_{\text{nodes}} \setminus G^{r \; \text{out}}_{\text{nodes}}  } 
  	\; \prod_{ i  \in {\cal U}  
		} E_{i \; i_{p}} \Big) 
\Bigg]
\\ 
{\hat{a}}_{i_{1},\ldots i_{k}}(G^{\prime}) &=
{\hat{a}}_{i_{1},\ldots i_{k}}(G^{\prime}_{\text{links}})
{\hat{a}}_{i_{1},\ldots i_{k}}(G^{\prime}_{\text{nodes}}) 
\\ &=
\Bigg [ \prod \limits_{(s^{\prime}, t^{\prime}) \in G^{\prime}_{\text{links}} } 
{\hat{a}}_{i_{s^{\prime}} i_{t^{\prime}}}
\Bigg] 
\Bigg[  \prod \limits_{v^{\prime} \in
G^{\prime}_{\text{nodes}} }  
{\hat{a}}_{i_{v^{\prime}} {\lambda^{\prime }}_{v^{\prime}}}\Bigg] 
 \\ 
  {{a}}_{i_{1},\ldots i_{k}}(G) & = 
{{a}}_{i_{1},\ldots i_{k}}(G_{\text{links}})
{{a}}_{i_{1},\ldots i_{k}}(G_{\text{nodes}}) 
\\ &=
\Bigg[ \prod \limits_{(s, t) \in G_{\text{links}} } 
a_{i_{s}
i_{t}}
\Bigg] 
\Bigg[  \prod \limits_{v \in G_{\text{nodes}} }  
a_{i_{v} \lambda_{v}} \Bigg] . %
\end{split}
\label{alt_eq2}
\end{equation}

\subsection{Index Set Notations}

In order to calculate operator products we introduce systematic index set notation as follows.

Define $L_\chi$, $L_\chi$, ${\cal L}_\chi$, ${\cal R}_\chi$,  for $\chi \in  \{1,2\}$:
\begin{equation}
\begin{split}
\text{lhs nodes}(r_1) \overset{\cal I}{\mapsto} {\cal I}(G^{1 \; \text{in}}_{\text{nodes}}) \equiv L_1 
	&  \quad \quad  \text{rhs nodes}(r_1) \overset{\cal I}{\mapsto} {\cal I}(G^{1 \; \text{out}}_{\text{nodes}}) \equiv R_1  \\
\text{lhs nodes}(r_2) \overset{\cal J}{\mapsto} {\cal J}(G^{2 \; \text{in}}_{\text{nodes}}) \equiv L_2 
	&  \quad \quad  \text{rhs nodes}(r_2) \overset{\cal J}{\mapsto} {\cal I}(G^{2 \; \text{out}}_{\text{nodes}}) \equiv  R_2  ; \\
\text{lhs links}(r_1) \overset{\cal I}{\mapsto} {\cal I}(G^{1 \; \text{in}}_{\text{links}}) \equiv {\cal L}_1 
	&  \quad \quad  \text{rhs links}(r_1) \overset{\cal I}{\mapsto} {\cal I}(G^{1 \; \text{out}}_{\text{links}}) \equiv  {\cal R}_1  \\
\text{lhs links}(r_2) \overset{\cal J}{\mapsto} {\cal J}(G^{2 \; \text{in}}_{\text{links}}) \equiv {\cal L}_2 
	&  \quad \quad  \text{rhs links}(r_2) \overset{\cal J}{\mapsto} {\cal I}(G^{2 \; \text{out}}_{\text{links}}) \equiv  {\cal R}_2  .
\end{split}
\end{equation}
In this notation the no-edge-cleanup semantics of Equation~(\ref{eq2}) becomes
\begin{equation}
\begin{split}
{\hat{W}}_{r_\chi} \; =&
{1 \over C_{r_\chi}(N_{\text{max free}})}
\rho _{r_\chi}({\lambdab}^{(\chi)}, {\lambdab^{\prime }}^{(\chi)}) 
\sum \limits_{{\cal I}_\chi  : L_\chi \cup R_\chi \overset{\text{1-1}}{\hookrightarrow} {\cal U}}
\\ & \times
  \Bigg [ \prod \limits_{(i_1, i_2) \in  {\cal R}_\chi} 
  	 {\hat{a}}_{i_1 i_2}
  \Bigg]  
  \Bigg [ \prod \limits_{i_5 \in  R_\chi }
  	  {\hat{a}}_{i_5 \, , \,  \lambda^{\prime {(1)}}_{{\cal I}^{-1}(i_5)}}
  \Bigg]  
  \Bigg [ \prod \limits_{(i_3, i_4) \in  {\cal L}_\chi} 
  	 { {a}}_{i_3 i_4}
  \Bigg]  
  \Bigg [ \prod \limits_{i_6 \in  L_\chi }
  	  {{a}}_{i_6 \, , \,  \lambda^{(1)}_{{\cal I}^{-1}(i_6)}}
  \Bigg]  
\end{split}
\label{equiv_rule_form}
\end{equation}
for $\chi \in \{1,2\}$, where ${\cal I}_{\chi=1} \equiv {\cal I}$ and ${\cal I}_{\chi=2} \equiv {\cal J}$.
Note that the middle square-bracketed terms commute trivially since elementary node and link
operators operate in different spaces.

Also in this notation, once again
\begin{equation}
\begin{split}
S &= \text{rhs}_1 \cap h^{-1}(\text{lhs}_2) & = G^{r_1 \; \text{out}}_{\text{nodes}} \cap h^{-1}(G^{r_2 \; \text{in}}_{\text{nodes}}) \\
h(S) &= \text{lhs}_2 \cap h(\text{rhs}_1) & = G^{r_2 \; \text{in}}_{\text{nodes}} \cap h^{-1}(G^{r_1 \; \text{out}}_{\text{nodes}}) \\
{\cal I}(S) & = {\cal J}(h(S)) = L_2 \cap R_1 \\
\overline{{\cal I}(S)} & = \overline{ L_2 \cap R_1 } = \overline{ L_2} \cup \overline{R_1 }   .
\end{split}
\end{equation}
Note also that 
\begin{equation}
{\cal L}_\chi \subseteq [ L_\chi \times L_\chi]$ and  ${\cal R}_\chi \subseteq [ R_\chi \times R_\chi]
\label{links_between_nodes}
\end{equation}
should be preserved inductively by rule-firing semantics.

Define ${\cal P}_\chi(i_1, i_2)$ = a predicate that determines which edges $E_{i_1, i_2}$ are hanging, if present, and should be deleted,
where $\chi \in \{1,2\}$.
It may be a predicate function:
${\cal P}_\chi[L_\chi, R_\chi, \ldots, G^{\chi \; \text{in}}_{\text{links}}, G^{\chi \; \text{out}}_{\text{links}}](i_1, i_2)$.
Also $P^T(i_1,i_2) \equiv P(i_2,i_1)$.
We will use one of several equivalent possibilities:
\begin{equation}
\begin{split}
{\cal P}_\chi &= [ ( L_\chi \setminus R_\chi ) \times {\cal U}] \; \\
\end{split}
\overset{\text{ duals }}{\longleftrightarrow}
\begin{split}
{\cal P}^*_\chi &= {\cal P}^T_\chi = [ {\cal U} \times  ( L_\chi \setminus R_\chi ) ] \\
\end{split}
\label{our_predicate}
\end{equation}
As before, ${\cal U}$ = the universe of possible node indices $i$.

(Any of these alternative formulations of ${\cal P}$ would be equivalent:
\begin{equation}
\begin{split}
\text {I}&: \,  \\
\text {II}&: \,  \\
 \text {III}&: \,  \\
 \text {IV}&:\, 
 \end{split}
\begin{split}
{\cal P}_\chi &= [ ( L_\chi \setminus R_\chi ) \times {\cal U}] \; \\
 {\cal P}_\chi &= [ ( L_\chi \setminus R_\chi ) \times \overline{L_\chi \cup R_\chi}] \; \\
{\cal P}_\chi &= [ ( L_\chi \setminus R_\chi ) \times {\cal U}] 
   \setminus  ({\cal L}_\chi \setminus {\cal R}_\chi ) ] \\
 {\cal P}_\chi &= [ ( L_\chi \setminus R_\chi ) \times \overline{L_\chi \cup R_\chi}] 
   \setminus  ({\cal L}_\chi \setminus {\cal R}_\chi ) ] ;
\end{split}
\overset{\text{ duals }}{\longleftrightarrow}
\begin{split}
{\cal P}^*_\chi &= {\cal P}^T_\chi = [ {\cal U} \times  ( L_\chi \setminus R_\chi ) ] \\
 {\cal P}^*_\chi &= {\cal P}^T_\chi = [ \overline{L_\chi \cup R_\chi} \times  ( L_\chi \setminus R_\chi ) ] \\
{\cal P}^*_\chi &= [  {\cal U} \times ( L_\chi \setminus R_\chi ) ] 
  \setminus ({\cal L}_\chi \setminus {\cal R}_\chi ) ] \\
{ \cal P}^*_\chi &= [ \overline{L_\chi \cup R_\chi} \times ( L_\chi \setminus R_\chi ) ] 
    \setminus ({\cal L}_\chi \setminus {\cal R}_\chi ) ] .
\end{split}
\end{equation}
But we will use Equation~(\ref{our_predicate}) = case I above, since it is the easiest to work with.)

Denote the sought-after ``compound rule'' for rules $r_1$ followed by $r_2$ as $r_{1;2}$.
Calculate
$L_{1;2}   = L_1 \cup (L_2 \setminus R_1)  =  L_1 \cup (L_2 \setminus {\cal I}(S))$
because  ${\cal I}(S) \subseteq R_1$ and $L_2 \cap R_1 \subseteq {\cal I}(S)$;
likewise
$R_{1;2} = R_2 \cup (R_1 \setminus L_2) = R_2 \cup (R_1 \setminus {\cal I}(S))$
because  ${\cal I}(S) \subseteq L_2$ and $  R_1 \cap L_2 \subseteq {\cal I}(S)$.
Similarly for ${\cal L}_{1;2}$ and ${\cal R}_{1;2}$.

Then we have these compound rule index set definitions:
\begin{equation}
\boxed{
\begin{split}
L_{1;2}  &= L_1 \cup (L_2 \setminus R_1) 
	& =  L_1 \cup (L_2 \setminus {\cal I}(S)) \\
R_{1;2} & = R_2 \cup (R_1 \setminus L_2) 
	& = R_2 \cup (R_1 \setminus {\cal I}(S)) \\
\Delta &= (L_2 \setminus R_2) \cap (R_1 \setminus L_1)  \; \equiv \; \overline{R_2} \cap L_2 \cap R_1 \cap \overline{L_1}   \\
{\cal L}_{1;2}  &= {\cal L}_1 \cup ({\cal L}_2 \setminus {\cal R}_1) 
	& =  {\cal L}_1 \cup ({\cal L}_2 \setminus {\cal I}(H_{1 \; \text{links}})) \\
{\cal R}_{1;2} & = {\cal R}_2 \cup ({\cal R}_1 \setminus {\cal L}_2) 
	& = {\cal R}_2 \cup ({\cal R}_1 \setminus {\cal I}(H_{1 \; \text{links}})) \\
{\cal D} &= ({\cal L}_2 \setminus {\cal R}_2) \cap ({\cal R}_1 \setminus {\cal L}_1) 
	 \; \equiv \; \overline{{\cal R}_2} \cap {\cal L}_2 \cap {\cal R}_1 \cap \overline{{\cal L}_1}   \\
\end{split}
} \; .
\label{set_defs}
\end{equation}
The index sets $\Delta$ and ${\cal D}$ above will turn up in the calculation of the next section.

From $a^2  = 0 = {\hat{a}}^2$, we have:
\begin{equation}
\begin{split}
(R_1 \setminus L_1) \cap (R_2 \setminus L_2) &= \varnothing \\
(L_1 \setminus R_1) \cap (L_2 \setminus R_2) &= \varnothing \\
({\cal R}_1 \setminus {\cal L}_1) \cap ({\cal R}_2 \setminus {\cal L}_2) &= \varnothing \\
({\cal L}_1 \setminus {\cal R}_1) \cap ({\cal L}_2 \setminus {\cal R}_2) &= \varnothing .
\end{split}
\end{equation}

\section{Calculations}
\label{sec_calculations}

\subsection{Commutation calculation - no edge cleanup}
\label{calculation-no-cleanup}

The product of two such operators is
(omitting for now the integral over parameters $X$)
\begin{equation}
\begin{split}
{\hat{W}}_{r_2} 
{\hat{W}}_{r_1} \; =
{1 \over C_{r_1}(N_{\text{max free}})}
  &
  {1 \over C_{r_2}(N_{\text{max free}})}
\rho _{r_1}({\lambdab}^{(1)}, {\lambdab^{\prime }}^{(1)}) 
\rho _{r_2}({\lambdab}^{(2)}, {\lambdab^{\prime }}^{(2)}) 
\\ & \times
\sum \limits_{{\cal J}  : L_2 \cup R_2 \overset{\text{1-1}}{\hookrightarrow} {\cal U}}
\quad 
\sum \limits_{{\cal I}  : L_1 \cup R_1 \overset{\text{1-1}}{\hookrightarrow} {\cal U}}
\\ & \times
  \Bigg [ \prod \limits_{(j_1, j_2) \in  {\cal R}_2 } 
  	 {\hat{a}}_{j_1 j_2}
  \Bigg]  
  \Bigg [ \prod \limits_{(j_3, j_4) \in  {\cal L}_2} 
  	 { {a}}_{j_3 j_4}
  \Bigg]  
  \Bigg [ \prod \limits_{j_5 \in  R_2 }
  	  {\hat{a}}_{j_5 \, , \,  \lambda^{\prime {(2)}}_{{\cal J}^{-1}(j_5)}}
  \Bigg]  
  \Bigg [ \prod \limits_{j_6 \in  L_2 }
  	  {{a}}_{j_6 \, , \,  \lambda^{(2)}_{{\cal J}^{-1}(j_6)}}
  \Bigg]  
\\ & \times
  \Bigg [ \prod \limits_{(i_1, i_2) \in  {\cal R}_1} 
  	 {\hat{a}}_{i_1 i_2}
  \Bigg]  
  \Bigg [ \prod \limits_{(i_3, i_4) \in  {\cal L}_1} 
  	 { {a}}_{i_3 i_4}
  \Bigg]  
  \Bigg [ \prod \limits_{i_5 \in  R_1 }
  	  {\hat{a}}_{i_5 \, , \,  \lambda^{\prime {(1)}}_{{\cal I}^{-1}(i_5)}}
  \Bigg]  
  \Bigg [ \prod \limits_{i_6 \in  L_1 }
  	  {{a}}_{i_6 \, , \,  \lambda^{(1)}_{{\cal I}^{-1}(i_6)}}
  \Bigg]  
\end{split}
\label{pred_prod_form_1a}
\end{equation}

Grouping the node operators together at the end, 
and grouping together terms that need to be commuted next
as $\big\{ \ldots \big\}_{\raisebox{.5pt}{\textcircled{\raisebox{-.9pt} {\small 1}}}}$ 
and $\big\{ \ldots \big\}_{\raisebox{.5pt}{\textcircled{\raisebox{-.9pt} {\small 2}}}}$,
this is:
\begin{equation}
\begin{split}
{\hat{W}}_{r_2} 
{\hat{W}}_{r_1} \; =
  &
{1 \over C_{r_1}(N_{\text{max free}})}
  {1 \over C_{r_2}(N_{\text{max free}})}
\rho _{r_1}({\lambdab}^{(1)}, {\lambdab^{\prime }}^{(1)}) 
\rho _{r_2}({\lambdab}^{(2)}, {\lambdab^{\prime }}^{(2)}) 
\\ & \times
\sum \limits_{{\cal J}  : L_2 \cup R_2 \overset{\text{1-1}}{\hookrightarrow} {\cal U}}
\sum \limits_{{\cal I}  : L_1 \cup R_1 \overset{\text{1-1}}{\hookrightarrow} {\cal U}}
  \Bigg [ \prod \limits_{(j_1, j_2) \in  {\cal R}_2 } 
  	 {\hat{a}}_{j_1 j_2}
  \Bigg]  
\\ & \times
\Bigg\{
  \Bigg [ \prod \limits_{(j_3, j_4) \in  {\cal L}_2} 
  	 { {a}}_{j_3 j_4}
  \Bigg]  
%
  \Bigg [ \prod \limits_{(i_1, i_2) \in  {\cal R}_1} 
  	 {\hat{a}}_{i_1 i_2}
  \Bigg]  
\Bigg\}_{\raisebox{.5pt}{\textcircled{\raisebox{-.9pt} {\small 1}}}}
  \Bigg [ \prod \limits_{(i_3, i_4) \in  {\cal L}_1} 
  	 { {a}}_{i_3 i_4}
  \Bigg]  
  \Bigg [ \prod \limits_{j_5 \in  R_2 }
  	  {\hat{a}}_{j_5 \, , \,  \lambda^{\prime {(2)}}_{{\cal J}^{-1}(j_5)}}
  \Bigg]  
 \\ & \times
\Bigg\{
  \Bigg [ \prod \limits_{j_6 \in  L_2 }
  	  {{a}}_{j_6 \, , \,  \lambda^{(2)}_{{\cal J}^{-1}(j_6)}}
  \Bigg]  
  \Bigg [ \prod \limits_{i_5 \in  R_1 }
  	  {\hat{a}}_{i_5 \, , \,  \lambda^{\prime {(1)}}_{{\cal I}^{-1}(i_5)}}
  \Bigg]  
\Bigg\}_{\raisebox{.5pt}{\textcircled{\raisebox{-.9pt} {\small 2}}}}
  \Bigg [ \prod \limits_{i_6 \in  L_1 }
  	  {{a}}_{i_6 \, , \,  \lambda^{(1)}_{{\cal I}^{-1}(i_6)}}
  \Bigg]  
\end{split}
\label{pred_prod_form_1b}
\end{equation}
Strategically rewriting the sum over ${\cal J}$, 
\begin{equation}
\boxed{
\begin{split}
{\hat{W}}_{r_2} 
{\hat{W}}_{r_1} \; =
  &
{1 \over C_{r_1}(N_{\text{max free}})}
  {1 \over C_{r_2}(N_{\text{max free}})}
\rho _{r_1}({\lambdab}^{(1)}, {\lambdab^{\prime }}^{(1)}) 
\rho _{r_2}({\lambdab}^{(2)}, {\lambdab^{\prime }}^{(2)}) 
\\ & \times
  \sum_{T \subseteq     G^{r_1 \; \text{in}}_{\text{nodes}}  \setminus G^{r_1 \; \text{out}}_{\text{nodes}} } \quad
  \sum_{\pi: T \overset{\text{\tiny 1--1}}{  \hookrightarrow } G^{r_2 \; \text{out}}_{\text{nodes}} \setminus G^{r_2 \; \text{in}}_{\text{nodes}}  } \quad
\sum_{S \subseteq G^{r_1 \; \text{out}}_{\text{nodes}} } \quad
  \sum_{\substack{
    {{{h}:S \overset{\text{\tiny 1--1}}{  \hookrightarrow } G^{r_2 \; \text{in}}_{\text{nodes}} } }
    }}
\\ &
\sum \limits_{{\cal I}  : L_1 \cup R_1 \overset{\text{1-1}}{\hookrightarrow} {\cal U}}
\quad
\sum \limits_{\substack{{\cal J}  : L_2 \cup R_2 \overset{\text{1-1}}{\hookrightarrow} {\cal U} \\ 
		\Im({\cal I}) \cap \Im({\cal J}) = {\cal I}(S) \cup {\cal I}(T) \ \\
		{\cal I}(S)={\cal J}(h(S)) \wedge {\cal I}(T)={\cal J}(\pi(T)) }}
  \Bigg [ \prod \limits_{(j_1, j_2) \in  {\cal R}_2} 
  	 {\hat{a}}_{j_1 j_2}
  \Bigg]  
\\ & \times
\Bigg\{
  \Bigg [ \prod \limits_{(j_3, j_4) \in  {\cal L}_2} 
  	 { {a}}_{j_3 j_4}
  \Bigg]  
  \Bigg [ \prod \limits_{(i_1, i_2) \in  {\cal R}_1} 
  	 {\hat{a}}_{i_1 i_2}
  \Bigg]  
%
\Bigg\}_{\raisebox{.5pt}{\textcircled{\raisebox{-.9pt} {\small 1}}}}
  \Bigg [ \prod \limits_{(i_3, i_4) \in  {\cal L}_1} 
  	 { {a}}_{i_3 i_4}
  \Bigg]  
  \Bigg [ \prod \limits_{j_5 \in  R_2 }
  	  {\hat{a}}_{j_5 \, , \,  \lambda^{\prime {(2)}}_{{\cal J}^{-1}(j_5)}}
  \Bigg]  
\\ & \times
\Bigg\{
  \Bigg [ \prod \limits_{j_6 \in  L_2 }
  	  {{a}}_{j_6 \, , \,  \lambda^{(2)}_{{\cal J}^{-1}(j_6)}}
  \Bigg]  
  \Bigg [ \prod \limits_{i_5 \in  R_1 }
  	  {\hat{a}}_{i_5 \, , \,  \lambda^{\prime {(1)}}_{{\cal I}^{-1}(i_5)}}
  \Bigg]  
\Bigg\}_{\raisebox{.5pt}{\textcircled{\raisebox{-.9pt} {\small 2}}}}
  \Bigg [ \prod \limits_{i_6 \in  L_1 }
  	  {{a}}_{i_6 \, , \,  \lambda^{(1)}_{{\cal I}^{-1}(i_6)}}
  \Bigg]  
\end{split}
}
\label{pred_prod_form_1c}
\end{equation}

In the controlled index model the sum over maps ${\cal J}$ will simplify
because $T$ and $\pi(T)$ will both be the null set.
In any case, note that
\begin{equation}
{\cal I}(T)  \subseteq L_1 \setminus R_1
\quad \text{ and } \quad
{\cal J}(\pi(T)) \subseteq R_2 \setminus L_2 \quad .
\label{T_bound}
\end{equation}

The 
more constraining
form to commute is
$\big\{ \ldots \big\}_{\raisebox{.5pt}{\textcircled{\raisebox{-.9pt} {\small 2}}}}$:
\begin{equation}
\begin{split}
\Bigg\{
  \Bigg [ \prod \limits_{j_6 \in  L_2 }
  	  {{a}}_{j_6 \, , \,  \lambda^{(2)}_{{\cal J}^{-1}(j_6)})}
  \Bigg]  
&
  \Bigg [ \prod \limits_{i_5 \in  R_1 }
  	  {\hat{a}}_{i_5 \, , \,  \lambda^{\prime {(1)}}_{{\cal I}^{-1}(i_5)})}
  \Bigg]  
\Bigg\}_{\raisebox{.5pt}{\textcircled{\raisebox{-.9pt} {\small 2}}}}
\\ & = 
  \Bigg [ \prod \limits_{j_{6  \; 0} \in  L_2 \cap {\cal I}(S) } \prod \limits_{j_{6 \; 1} \in  L_2 \cap \overline{{\cal I}(S)} }
  	  {{a}}_{j_{6  \; 0} \, , \,  \lambda^{(2)}_{{\cal J}^{-1}(j_{6  \; 0})}}
  	  {{a}}_{j_{6  \; 1} \, , \,  \lambda^{(2)}_{{\cal J}^{-1}(j_{6  \; 1})}}
  \Bigg]  
\\ & \quad \times
  \Bigg [ \prod \limits_{i_{5  \; 0} \in  R_1 \cap {\cal I}(S) } \prod \limits_{i_{5 \; 1} \in  R_1 \cap \overline{{\cal I}(S)} }
  	  {\hat{a}}_{i_{5  \; 0} \, , \,  \lambda^{\prime {(1)}}_{{\cal I}^{-1}(i_{5  \; 0})}}
  	  {\hat{a}}_{i_{5  \; 1} \, , \,  \lambda^{\prime {(1)}}_{{\cal I}^{-1}(i_{5  \; 1})}}
  \Bigg]  .
\end{split}
\label{pred_prod_subform_2a}
\end{equation}

Using Equation~(\ref{key_comm})
$a_{j \, , \, \lambda} \hat{a}_{i \, , \, \lambda^{\prime}} 
= (1-\delta_{i j} ) \hat{a}_{i \, , \, \lambda^{\prime}} a_{j \, , \, \lambda} +\delta_{i j}  \delta_{\lambda \lambda^{\prime}}Y_{j \, , \, \lambda^{\prime}}$,
so that only indices in ${\cal I}(S)$ may fail to commute,
this becomes
$\big\{ \ldots \big\}_{\raisebox{.5pt}{\textcircled{\raisebox{-.9pt} {\small 2}}}}$ :
\begin{equation}
\begin{split}
\Bigg\{
  \Bigg [ \prod \limits_{j_6 \in  L_2 }
  	  {{a}}_{j_6 \, , \,  \lambda^{(2)}_{{\cal J}^{-1}(j_6)})}
  \Bigg]  
&
  \Bigg [ \prod \limits_{i_5 \in  R_1 }
  	  {\hat{a}}_{i_5 \, , \,  \lambda^{\prime {(1)}}_{{\cal I}^{-1}(i_5)})}
  \Bigg]  
\Bigg\}_{\raisebox{.5pt}{\textcircled{\raisebox{-.9pt} {\small 2}}}}
\\ & = 
  { \prod \limits_{j_{6  \; 0} \in  L_2 \cap{\cal I}(S) = {\cal I}(S) = L_2 \cap R_1} } 
  	\quad { \prod \limits_{j_{6 \; 1} \in  L_2 \cap \overline{{\cal I}(S)} = L_2 \cap \overline{R_1} } } \quad
  { \prod \limits_{i_{5  \; 0} \in  R_1 \cap {\cal I}(S) }  } \quad { \prod \limits_{i_{5 \; 1} \in  R_1 \cap \overline{{\cal I}(S)} } } 
\\ & \quad 
  	  {{a}}_{j_{6  \; 0} \, , \,  \lambda^{(2)}_{{\cal J}^{-1}(j_{6  \; 0})}}
  	  {{a}}_{j_{6  \; 1} \, , \,  \lambda^{(2)}_{{\cal J}^{-1}(j_{6  \; 1})}}
  	  {\hat{a}}_{i_{5  \; 0} \, , \,  \lambda^{\prime {(1)}}_{{\cal I}^{-1}(i_{5  \; 0})}}
  	  {\hat{a}}_{i_{5  \; 1} \, , \,  \lambda^{\prime {(1)}}_{{\cal I}^{-1}(i_{5  \; 1})}}
\\ & = 
  { \prod \limits_{j_{6  \; 0} \in  L_2 \cap R_1 }  } 
  	\quad { \prod \limits_{j_{6 \; 1} \in  L_2 \cap \overline{R_1} } } \quad
  { \prod \limits_{i_{5  \; 0} \in  L_2 \cap R_1}  } \quad { \prod \limits_{i_{5 \; 1} \in  R_1 \cap \overline{L_2} } } 
\\ & \quad 
  	  {\hat{a}}_{i_{5  \; 1} \, , \,  \lambda^{\prime {(1)}}_{{\cal I}^{-1}(i_{5  \; 1})}}
  	  {{a}}_{j_{6  \; 1} \, , \,  \lambda^{(2)}_{{\cal J}^{-1}(j_{6  \; 1})}}
  	  {{a}}_{j_{6  \; 0} \, , \,  \lambda^{(2)}_{{\cal J}^{-1}(j_{6  \; 0})}}
  	  {\hat{a}}_{i_{5  \; 0} \, , \,  \lambda^{\prime {(1)}}_{{\cal I}^{-1}(i_{5  \; 0})}}
\\ & = 
  { \prod \limits_{j_{6 \; 1} \in  L_2 \cap \overline{R_1} } } \quad
  { \prod \limits_{i_{5  \; 0} \in  L_2 \cap R_1}  } \quad { \prod \limits_{i_{5 \; 1} \in  R_1 \cap \overline{L_2} } } 
\\ & \quad 
  	  {\hat{a}}_{i_{5  \; 1} \, , \,  \lambda^{\prime {(1)}}_{{\cal I}^{-1}(i_{5  \; 1})}}
  	  {{a}}_{j_{6  \; 1} \, , \,  \lambda^{(2)}_{{\cal J}^{-1}(j_{6  \; 1})}}
	  	\delta_{\lambda^{\prime {(1)}}_{{\cal I}^{-1}(i_{5  \; 0})}  \; , \; \lambda^{(2)}_{{\cal J}^{-1}(j_{6  \; 1}) = h({\cal I}^{-1}(i_{5  \; 0})) } }
  	  Y_{i_{5  \; 0} \, , \,  \lambda^{\prime {(1)}}_{{\cal I}^{-1}(i_{5  \; 0})}}
\end{split}
\label{pred_prod_subform_2b}
\end{equation}
and thus
\begin{equation}
\boxed{
\begin{split}
\Bigg\{
  \Bigg [ \prod \limits_{j_6 \in  L_2 }
  	  {{a}}_{j_6 \, , \,  \lambda^{(2)}_{{\cal J}^{-1}(j_6)})}
  \Bigg]  
&
  \Bigg [ \prod \limits_{i_5 \in  R_1 }
  	  {\hat{a}}_{i_5 \, , \,  \lambda^{\prime {(1)}}_{{\cal I}^{-1}(i_5)})}
  \Bigg]  
\Bigg\}_{\raisebox{.5pt}{\textcircled{\raisebox{-.9pt} {\small 2}}}}
\\ & = 
    \Bigg [ 
  \prod \limits_{i_{5 \; 1} \in  R_1  \setminus L_2 } 
  	  {\hat{a}}_{i_{5  \; 1} \, , \,  \lambda^{\prime {(1)}}_{{\cal I}^{-1}(i_{5  \; 1})}}
 \Bigg] \quad \Bigg[
	  \prod \limits_{j_{6 \; 1} \in  L_2 \setminus R_1 } 
  	  {{a}}_{j_{6  \; 1} \, , \,  \lambda^{(2)}_{{\cal J}^{-1}(j_{6  \; 1})}}
  \Bigg]  
\\ & \quad \times
    \Bigg [ \prod \limits_{i_{5  \; 0} \in  L_2 \cap R_1 } 
	  	\delta_{\lambda^{\prime {(1)}}_{{\cal I}^{-1}(i_{5  \; 0})}  \; , \; \lambda^{(2)}_{ h({\cal I}^{-1}(i_{5  \; 0})) } }
  	  Y_{i_{5  \; 0} \, , \,  \lambda^{\prime {(1)}}_{{\cal I}^{-1}(i_{5  \; 0})}}
  \Bigg]  
\end{split}
}
\label{pred_prod_subform_2c}
\end{equation}
since on $S$, ${\cal J}^{-1} = h \circ {\cal I}^{-1}$.
The last line implements label-checking in the node correspondence
portion of graph matching between 
a subgraph $H(S,h)$ of the output graph of rule $r_1$ 
and a corresponding subgraph of the input graph of rule $r_2$.

We must now simplify
 $\Big\{
 \ldots
 \Big\}_{\raisebox{.5pt}{\textcircled{\raisebox{-.9pt} {\small 1}}}}
$
by commuting its leftmost factor,
$ \Big [ \prod \limits_{(j_3, j_4) \in {\cal R}_2 }{ {a}}_{j_3 j_4} \Big]  $,
to the right of its rightmost factor,
$  \Big [ \prod \limits_{(i_1, i_2) \in  {\cal R}_1}  {\hat{a}}_{i_1 i_2} \Big]  $.
To this end, using Equation~(\ref{binary_algebra_1})
and the conditions $\text{Im}({\cal I}) \cap \text{Im}({\cal J}) = {\cal I}(S) \cup {\cal I}(T)$,
and (from Equation (\ref{T_bound}) )
${\cal I}(T) \cap R_1 = \varnothing  \wedge {\cal J}(\pi(T)) \cap L_2 = \varnothing$,
we compute:
\begin{equation}
\begin{split}
 & \Bigg\{
   \Bigg [ \prod \limits_{(j_3, j_4) \in  {\cal L}_2 
   	 }
  	 { {a}}_{j_3 j_4}
  \Bigg] 
 \Bigg [ 
 \prod \limits_{(i_1, i_2) \in {\cal R}_1 }
  	 {\hat{a}}_{i_1 i_2}
  \Bigg]  
\Bigg\}_{\raisebox{.5pt}{\textcircled{\raisebox{-.9pt} {\small 1}}}}
= 
 \Bigg [ 
   \prod \limits_{(j_3, j_4) \in  {\cal L}_2 
   	}
 \quad
 \prod \limits_{(i_1, i_2) \in {\cal R}_1 }
  	 { {a}}_{j_3 j_4}
  	 {\hat{a}}_{i_1 i_2}
  \Bigg]  
\\ & = 
 \Bigg [ 
   \prod \limits_{(i_1, i_2)  \in  {\cal R}_1
    \setminus  {\cal L}_2 
		\subseteq \overline {[L_2 \cap R_1 \; \times \; L_2 \cap R_1 ]}
   	 } 
  	 {\hat{a}}_{i_1 i_2}
  \Bigg]   \quad \Bigg [ 
   \prod \limits_{(j_3, j_4) \in 
   \begin{split}
   	 & ({\cal L}_2 
		\subseteq \overline {[L_2 \cap R_1 \; \times \; L_2 \cap R_1 ]} ) 
	 \setminus {\cal R}_1 
   \end{split} }
  	 { {a}}_{j_3 j_4}
  \Bigg]  
 \\ & \quad \times 
 \Bigg [ 
   \prod \limits_{(j_7, j_8) \in 
   \begin{split}
   	 & {\cal L}_2 
	 \cap  {\cal R}_1 
	 \subseteq [L_2 \cap R_1 \; \times \; L_2 \cap R_1 ]
   \end{split} }
  	 Z_{j_7 j_8}
  \Bigg]  
\end{split}
\label{graph_match_2}
\end{equation}

Thus in our case,
\begin{equation}
\begin{split}
 &
 \Bigg\{
   \Bigg [ \prod \limits_{(j_3, j_4) \in  {\cal L}_2 
   		}
  	 { {a}}_{j_3 j_4}
  \Bigg] 
 \Bigg [ 
 \prod \limits_{(i_1, i_2) \in {\cal R}_1} 
  	 {\hat{a}}_{i_1 i_2}
  \Bigg]  
\Bigg\}_{\raisebox{.5pt}{\textcircled{\raisebox{-.9pt} {\small 1}}}}
 \\ & \quad = 
 \Bigg [ 
 \prod \limits_{(i_1, i_2) \in
      {\cal R}_1 
      \setminus  {\cal L}_2 
	}
  	 {\hat{a}}_{i_1 i_2}
  \Bigg]   \quad \Bigg [ 
   \prod \limits_{(j_3, j_4) \in 
   	 {\cal L}_2 
	 	}
  	 { {a}}_{j_3 j_4}
  \Bigg]  
\quad 
 \Bigg [ 
   \prod \limits_{(j_7, j_8) \in 
   	 {\cal L}_2 
	 \cap   {\cal R}_1 
	\equiv {\cal I} (H_{\text{links}} ) 
   }
  	 Z_{j_7 j_8}
  \Bigg]  
\end{split}
\label{graph_match_3a}
\end{equation}
Here 
${\cal I} (H_{\text{links}}) \equiv 
   	 {\cal L}_2 
	 \cap  {\cal R}_1 
	 $, 
or equivalently
$H_{\text{links}} \equiv 
	 G^{r_1 \; \text{out}}_{\text{links}}
	 \cap   
   	 {\cal I}^{-1}( {\cal J}(G^{r_2 \; \text{in}}_{\text{links}}) )
	 = G^{r_1 \; \text{out}}_{\text{links}} \cap   
   	 h^{-1}(G^{r_2 \; \text{in}}_{\text{links}})  
$;
likewise
$H_{\text{nodes}} \equiv 
	 G^{r_1 \; \text{out}}_{\text{nodes}}
	 \cap   
   	 h^{-1}(G^{r_2 \; \text{in}}_{\text{nodes}})  $.

Thus we have:

\begin{lemma}
$H(S,h)$ must be the {\rm maximal} common 
subgraph of both $G^{r_1 \; \text{out}}$ and $G^{r_2 \; \text{in}}$,
for any given choice of nodes $S$ in $G^{r_1 \; \text{out}}$ and
1-1 corresponding nodes $h(S)$ in  $G^{r_2 \; \text{in}}$.
From factor ${\raisebox{.5pt}{\textcircled{\raisebox{-.9pt} {\small 2}}}}$ we
can restrict $S$ to sets of nodes whose labels match
in $G^{r_2 \; \text{in}}_{\text{nodes}}$ and $G^{r_1 \; \text{out}}_{\text{nodes}}$.
For any such $H$, we can commute the link operators as follows:
\begin{equation}
\boxed{
\begin{split}
 &
   \Bigg [ \prod \limits_{(j_3, j_4) \in  {\cal L}_2 
   	}
  	 { {a}}_{j_3 j_4}
  \Bigg] 
 \Bigg [ 
 \prod \limits_{(i_1, i_2) \in {\cal R}_1 
 	}
  	 {\hat{a}}_{i_1 i_2}
  \Bigg]  
 \\ & \quad = 
 \Bigg [ 
 \prod \limits_{(i_1, i_2) \in
      {\cal I}(G^{r_1 \; \text{out}}_{\text{links}} \setminus   H_{\text{links}} ) }
  	 {\hat{a}}_{i_1 i_2}
  \Bigg]   \quad \Bigg [ 
   \prod \limits_{(j_3, j_4) \in 
   	 {\cal J} (G^{r_2 \; \text{in}}_{\text{links}}  \setminus h^{-1}( H_{\text{links}} ) )  
	 }
  	 { {a}}_{j_3 j_4}
  \Bigg]  
\quad 
 \Bigg [ 
   \prod \limits_{(j_7, j_8) \in 
   	{\cal I} (H_{\text{links}}) \equiv 
   	 {\cal L}_2  
	 \cap   {\cal R}_1 }
  	 Z_{j_7 j_8}
  \Bigg]  
\end{split}
}
\label{graph_match_3b}
\end{equation}
The last factor above
augments the graph matching of Equation~(\ref{pred_prod_subform_2c})
by implementing the edge-checking or link correspondence
portion of graph matching between 
a subgraph $H(S,h)$ of the output graph of rule $r_1$ 
and a corresponding subgraph of the input graph of rule $r_2$.
\end{lemma}

No proper subsets of ${\cal L}_2 \cap {\cal R}_1$ from commuting 
creation and annihilation operators need to be considered,
because the $Z$ factor in the last term,
arising from Equation~(\ref{pred_prod_form_1c}),
is already a sum of
two terms: $I$ and $-N$
corresponding to commutation with index miss and hit.

Thus, the ``Common$(G_1,G_2)$'' set of shared subgraphs $H$ that we sum over in
the graph rewrite commutator is:

{\bf Definition} Common$(G_1,G_2)$ = An isomorphic pair of 
(edge-maximal) labelled subgraphs 
$H_1 \simeq H_2 \simeq H$,
with graph embeddings $H_1 \hookrightarrow G_1$ and $H_2 \hookrightarrow G_2$.

From ${\cal I}(S) = {\cal I}((H_{\text{nodes}}) = L_2 \cap R_1$ and edge-maximality
we conclude ${\cal I}(H_{\text{links}}) = {\cal L}_2 \cap {\cal R}_1$, whence
Equation (\ref{graph_match_3b}) becomes
\begin{equation}
\boxed{
\begin{split}
 &
   \Bigg [ \prod \limits_{(j_3, j_4) \in  {\cal L}_2 
   	}
  	 { {a}}_{j_3 j_4}
  \Bigg] 
 \Bigg [ 
 \prod \limits_{(i_1, i_2) \in {\cal R}_1 
 	}
  	 {\hat{a}}_{i_1 i_2}
  \Bigg]  
 \\ & \quad = 
 \Bigg [ 
 \prod \limits_{(i_1, i_2) \in 
 	{\cal L}_2 \setminus {\cal R}_1 }
  	 {\hat{a}}_{i_1 i_2}
  \Bigg]   \quad \Bigg [ 
   \prod \limits_{(j_3, j_4) \in 
   	{\cal R}_1 \setminus {\cal L}_2
	 }
  	 { {a}}_{j_3 j_4}
  \Bigg]  
\quad 
 \Bigg [ 
   \prod \limits_{(j_7, j_8) \in 
   	 {\cal L}_2  
	 \cap   {\cal R}_1 }
  	 Z_{j_7 j_8}
  \Bigg]  
\end{split}
}
\label{graph_match_4}
\end{equation}


We now assemble partial results of
Equations~(\ref{pred_prod_form_1c}), (\ref{pred_prod_subform_2c}), 
and (\ref{graph_match_4}):
\begin{equation}
\begin{split}
{\hat{W}}_{r_2} 
{\hat{W}}_{r_1} \; =
  &
{1 \over C_{r_1}(N_{\text{max free}})}
  {1 \over C_{r_2}(N_{\text{max free}})}
\rho _{r_1}({\lambdab}^{(1)}, {\lambdab^{\prime }}^{(1)}) 
\rho _{r_2}({\lambdab}^{(2)}, {\lambdab^{\prime }}^{(2)}) 
\\ & \times
  \sum_{T \subseteq     G^{r_1 \; \text{in}}_{\text{nodes}}  \setminus G^{r_1 \; \text{out}}_{\text{nodes}} } \quad
  \sum_{\pi: T \overset{\text{\tiny 1--1}}{  \hookrightarrow } G^{r_2 \; \text{out}}_{\text{nodes}} \setminus G^{r_2 \; \text{in}}_{\text{nodes}}  } \quad
\sum_{S \subseteq G^{r_1 \; \text{out}}_{\text{nodes}} } \quad
  \sum_{\substack{
    {{{h}:S \overset{\text{\tiny 1--1}}{  \hookrightarrow } G^{r_2 \; \text{in}}_{\text{nodes}} } }
    }}
\\ &
\sum \limits_{{\cal I}  : L_1 \cup R_1 \overset{\text{1-1}}{\hookrightarrow} {\cal U}}
\quad
\sum \limits_{\substack{{\cal J}  : L_2 \cup R_2 \overset{\text{1-1}}{\hookrightarrow} {\cal U} \\ 
		\Im({\cal I}) \cap \Im({\cal J}) = {\cal I}(S) \cup {\cal I}(T) \ \\
		{\cal I}(S)={\cal J}(h(S)) = L_2 \cap R_1 \; \wedge \; {\cal I}(T)={\cal J}(\pi(T)) }}
\Bigg [ 
  { \prod \limits_{(j_{1 }, j_{2}) \in {\cal R}_2 }}
  	 {\hat{a}}_{(j_{1}, j_{2 })}
 \Bigg] 
\\ & \quad 
\Bigg\{
\Bigg [ 
 \prod \limits_{(i_1, i_2) \in
 	{\cal R}_1 \setminus {\cal L}_2 }
  	 {\hat{a}}_{i_1 i_2}
  \Bigg]   \quad \Bigg [ 
   \prod \limits_{(j_3, j_4) \in 
   	{\cal L}_2 \setminus {\cal R}_1
	  }
  	 { {a}}_{j_3 j_4}
  \Bigg]  
 \Bigg [ 
   \prod \limits_{(j_7, j_8) \in 
   	{\cal I} (H_{\text{links}}) \equiv 
   	  {\cal L}_2 
	 \cap  {\cal R}_1 
	 }
  	 Z_{j_7 j_8}
  \Bigg]  
\Bigg\}_{\raisebox{.5pt}{\textcircled{\raisebox{-.9pt} {\small 1}}}}
 \\ & \quad \times 
  \Bigg [ \prod \limits_{(i_3, i_4) \in {\cal L}_1 
  	}
  	 { {a}}_{i_3 i_4}
  \Bigg]  
  \Bigg [ \prod \limits_{j_5 \in  R_2 }
  	  {\hat{a}}_{j_5 \, , \,  \lambda^{\prime {(2)}}_{{\cal J}^{-1}(j_5)}}
  \Bigg]  
\\ & \times
\Bigg\{
   \Bigg[  \prod \limits_{i_{5 \; 1} \in  R_1 \setminus L_2 } 
  	  {\hat{a}}_{i_{5  \; 1} \, , \,  \lambda^{\prime {(1)}}_{{\cal I}^{-1}(i_{5  \; 1})}}
    \Bigg]\quad  \Bigg[ \prod \limits_{j_{6 \; 1} \in  L_2 \setminus R_1 } 
  	  {{a}}_{j_{6  \; 1} \, , \,  \lambda^{(2)}_{{\cal J}^{-1}(j_{6  \; 1})}}
  \Bigg]  
\\ &  \times
    \Bigg [ \prod \limits_{i_{5  \; 0} \in  R_1 \cap {\cal I}(S) = {\cal I}(S) } 
	  	\delta_{\lambda^{\prime {(1)}}_{{\cal I}^{-1}(i_{5  \; 0})}  \; , \; \lambda^{(2)}_{ h({\cal I}^{-1}(i_{5  \; 0})) } }
  \Bigg]  \text{   (a commuting scalar)}
\\ & \quad \times
     \Bigg [ \prod \limits_{i_{5  \; 0} \in  R_1 \cap {\cal I}(S) = {\cal I}(S) } 
 	  Y_{i_{5  \; 0} \, , \,  \lambda^{\prime {(1)}}_{{\cal I}^{-1}(i_{5  \; 0})}}
  \Bigg]  
\Bigg\}_{\raisebox{.5pt}{\textcircled{\raisebox{-.9pt} {\small 2}}}}
  \Bigg [ \prod \limits_{i_6 \in  L_1 }
  	  {{a}}_{i_6 \, , \,  \lambda^{(1)}_{{\cal I}^{-1}(i_6)}}
  \Bigg]  
\end{split}
\label{pred_prod_form_11}
\end{equation}

Ungrouping 
$\big\{ \ldots \big\}_{\raisebox{.5pt}{\textcircled{\raisebox{-.9pt} {\small 1}}}}$ and 
$\big\{ \ldots \big\}_{\raisebox{.5pt}{\textcircled{\raisebox{-.9pt} {\small 2}}}}$,
and using 
the identities
$Z_\alpha a_\alpha = a_\alpha$, $Y_\alpha a_\alpha = a_\alpha$, 
and $(a_\alpha)^2=0= (\hat{a}_\alpha)^2$, and regrouping, 
\begin{equation}
\boxed{
\begin{split}
{\hat{W}}_{r_2} 
{\hat{W}}_{r_1} \; =
  &
{1 \over C_{r_1}(N_{\text{max free}})}
  {1 \over C_{r_2}(N_{\text{max free}})}
\rho _{r_1}({\lambdab}^{(1)}, {\lambdab^{\prime }}^{(1)}) 
\rho _{r_2}({\lambdab}^{(2)}, {\lambdab^{\prime }}^{(2)}) 
\\ & \times
  \sum_{T \subseteq     G^{r_1 \; \text{in}}_{\text{nodes}}  \setminus G^{r_1 \; \text{out}}_{\text{nodes}} } \quad
  \sum_{\pi: T \overset{\text{\tiny 1--1}}{  \hookrightarrow } G^{r_2 \; \text{out}}_{\text{nodes}} \setminus G^{r_2 \; \text{in}}_{\text{nodes}}  } \quad
 \\ & \times
\sum_{S \subseteq G^{r_1 \; \text{out}}_{\text{nodes}} } \quad
  \sum_{\substack{
    {{{h}:S \overset{\text{\tiny 1--1}}{  \hookrightarrow } G^{r_2 \; \text{in}}_{\text{nodes}} } }
    }}
    \Bigg [ \prod \limits_{i_{5   \; 0} \in   L_2 \cap R_1 = {\cal I}(S)  } 
	  	\delta_{\lambda^{\prime {(1)}}_{{\cal I}^{-1}(i_{5  \; 0})}  \; , \; \lambda^{(2)}_{ h({\cal I}^{-1}(i_{5  \; 0})) } }
    \Bigg]  
    \quad  \text{ // defines $\sum_H$ and adjusts $1 \over{C}$ }
\\ &
\sum \limits_{{\cal I}  : L_1 \cup R_1 \overset{\text{1-1}}{\hookrightarrow} {\cal U}}
\quad
\sum \limits_{\substack{{\cal J}  : L_2 \cup R_2 \overset{\text{1-1}}{\hookrightarrow} {\cal U} \\ 
		\Im({\cal I}) \cap \Im({\cal J}) = {\cal I}(S) \cup {\cal I}(T) \ \\
		{\cal I}(S)={\cal J}(h(S)) = L_2 \cap R_1 \; \wedge \; {\cal I}(T)={\cal J}(\pi(T)) }}
\quad 
\Bigg\{
\Bigg [ 
  { \prod \limits_{(j_{1 }, j_{2}) \in {\cal R}_2 }}
  	 {\hat{a}}_{(j_{1}, j_{2 })}
 \Bigg] 
\Bigg [ 
 \prod \limits_{(i_1, i_2) \in
 	{\cal R}_1 \setminus {\cal L}_2 }
  	 {\hat{a}}_{i_1 i_2}
  \Bigg]  \quad
\Bigg\}_{\raisebox{.5pt}{\textcircled{\raisebox{-.9pt} {\small 3}}}}
\\ & \quad \times
\Bigg\{
  \Bigg [ \prod \limits_{j_5 \in  
  	R_2 \setminus (R_1 \cap \overline{L_2 \cap R_1} ) 
	= R_2 \setminus (R_1 \setminus L_2) 
	}
  	  {\hat{a}}_{j_5 \, , \,  \lambda^{\prime {(2)}}_{{\cal J}^{-1}(j_5)}}
  \Bigg]  
\Bigg [ 
  \prod \limits_{i_{5 \; 1} \in  R_1 \setminus L_2 
   		= {\cal I}(   G^{r_1 \; \text{out}}_{\text{nodes}}   \setminus S)}
  	  {\hat{a}}_{i_{5  \; 1} \, , \,  \lambda^{\prime {(1)}}_{{\cal I}^{-1}(i_{5  \; 1})}}
  \Bigg]  
\Bigg\}_{\raisebox{.5pt}{\textcircled{\raisebox{-.9pt} {\small 4}}}}
\\ &  \times
\Bigg\{
 \Bigg [ 
   \prod \limits_{(j_3, j_4) \in 
   	{\cal L}_2 \setminus {\cal R}_1
	 }
  	 { {a}}_{j_3 j_4}
  \Bigg]  
  \Bigg [ \prod \limits_{(i_3, i_4) \in  {\cal L}_1 } 
  	 { {a}}_{i_3 i_4}
  \Bigg]  
  \Bigg\}_{\raisebox{.5pt}{\textcircled{\raisebox{-.9pt} {\small 5}}}}
\\ & \quad  \times
\Bigg\{
    \Bigg [ \prod \limits_{j_{6 \; 1} \in  L_2 \setminus R_1 
        =  {\cal J} (G^{r_2 \; \text{in}}_{\text{nodes}}  \setminus h^{-1}(S) ) } \quad
  	  {{a}}_{j_{6  \; 1} \, , \,  \lambda^{(2)}_{{\cal J}^{-1}(j_{6  \; 1})}}
  \Bigg]  
  \Bigg [ \prod \limits_{i_6 \in  L_1 } 
  	  {{a}}_{i_6 \, , \,  \lambda^{(1)}_{{\cal I}^{-1}(i_6)}}
  \Bigg]  
\Bigg\}_{\raisebox{.5pt}{\textcircled{\raisebox{-.9pt} {\small  6}}}}
\\ & \times
\Bigg\{
 \Bigg [ 
   \prod \limits_{(j_7, j_8) \in 
   	{\cal I} (H_{\text{links}}) \setminus {\cal L}_1 \equiv 
   	 ({\cal L}_2 
	 \cap  {\cal R}_1 )
	  \setminus {\cal L}_1}
  	 Z_{j_7 j_8}
  \Bigg]  
    \Bigg [ \prod \limits_{i_{5  \; 0 \; 0} \in   {\cal I}(H_{\text{nodes}})\setminus L_1  = (L_2 \cap R_1 ) \setminus L_1 } 
  	  Y_{i_{5  \; 0 \; 0} \, , \,  \lambda^{\prime {(1)}}_{{\cal I}^{-1}(i_{5  \; 0 \; 0})}}
  \Bigg]  
\Bigg\}_{\raisebox{.5pt}{\textcircled{\raisebox{-.9pt} {\small 7}}}}
\end{split}
}
\label{pred_prod_form_12}
\end{equation}


Then, using
Equation~(\ref{set_defs}) to redefine the index set domains,
along with extended indices $i_n^{\star}$ with $\star$ superscripts to run over them
(according to the index map
${\cal I}^{\star}$ which extends ${\cal I}$ with nonoverlapping assignments from ${\cal J}$ as appropriate);
and using
 the index allocation scheme to force $T=\varnothing$;
 we have
\begin{equation}
\boxed{
\begin{split}
{\hat{W}}_{r_2} 
{\hat{W}}_{r_1} \; =
  &
{1 \over C_{r_1}(N_{\text{max free}})}
  {1 \over C_{r_2}(N_{\text{max free}})}
\rho _{r_1}({\lambdab}^{(1)}, {\lambdab^{\prime }}^{(1)}) 
\rho _{r_2}({\lambdab}^{(2)}, {\lambdab^{\prime }}^{(2)}) 
\sum_{S \subseteq G^{r_1 \; \text{out}}_{\text{nodes}} } \quad
  \sum_{\substack{
    {{{h}:S \overset{\text{\tiny 1--1}}{  \hookrightarrow } G^{r_2 \; \text{in}}_{\text{nodes}} } }
    }}
\\ &
\sum \limits_{{\cal I}  : L_1 \cup R_1 \overset{\text{1-1}}{\hookrightarrow} {\cal U}}
\quad
\sum \limits_{\substack{{\cal J}  : L_2 \cup R_2 \overset{\text{1-1}}{\hookrightarrow} {\cal U} \\ 
		\Im({\cal I}) \cap \Im({\cal J}) = {\cal I}(S) \\
		{\cal I}(S)={\cal J}(h(S)) = L_2 \cap R_1}}
    \Bigg [ \prod \limits_{i_{5   \; 0} \in   L_2 \cap R_1 = {\cal I}(S)  } 
	  	\delta_{\lambda^{\prime {(1)}}_{{\cal I}^{-1}(i_{5  \; 0})}  \; , \; \lambda^{(2)}_{ h({\cal I}^{-1}(i_{5  \; 0})) } }
    \Bigg]  
\\ & \quad
\Bigg [ 
 \prod \limits_{(i^{\star}_1, i^{\star}_2) \in
 	{\cal R}_{1;2}  = {\cal R}_2 \cup ({\cal R}_1 \setminus {\cal L}_2 ) }
  	 {\hat{a}}_{i_1 i_2}
  \Bigg]_{\raisebox{.5pt}{\textcircled{\raisebox{-.9pt} {\small 3}}}}
\Bigg [   \prod \limits_{i^{\star}_1 \in  R_{1;2} }
  	  {\hat{a}}_{i^{\star}_1 \, , \,  \lambda^{\prime {(1;2)}}_{{\cal I}^{\star -1}_1(i^{\star})}}
  \Bigg] _{\raisebox{.5pt}{\textcircled{\raisebox{-.9pt} {\small 4}}}}
\\ & \quad \times
\Bigg [ \prod \limits_{(i^{\star}_3, i^{\star}_4) \in  {\cal L}_{1;2}  =  {\cal L}_1  \cup ({\cal L}_2 \setminus {\cal R}_1)}
  	 { {a}}_{i_3 i_4}
  \Bigg]_{\raisebox{.5pt}{\textcircled{\raisebox{-.9pt} {\small 5}}}}
  \Bigg [ \prod \limits_{i^{\star}_2 \in  L_{1;2} } 
  	  {{a}}_{i^{\star}_2 \, , \,  \lambda^{(1;2)}_{{\cal I}^{\star -1}(i^{\star}_2 )}}
  \Bigg]  
_{\raisebox{.5pt}{\textcircled{\raisebox{-.9pt} {\small 6}}}}
\\ & \times
\Bigg\{
 \Bigg [ 
   \prod \limits_{(j_7, j_8) \in {\cal D} =
   	 ({\cal L}_2 
	 \cap  {\cal R}_1 )
	  \setminus {\cal L}_1 \setminus {\cal R}_{2} }
  	 Z_{j_7 j_8}
  \Bigg]  
    \Bigg [ \prod \limits_{i_{5  \; 0 \; 0} \in  \Delta = (L_2 \cap R_1 ) \setminus L_1 \setminus R_{2}  } 
  	  Y_{i_{5  \; 0 \; 0} \, , \,  \lambda^{\prime {(1)}}_{{\cal I}^{-1}(i_{5  \; 0 \; 0})}}
  \Bigg]  
\Bigg\}_{\raisebox{.5pt}{\textcircled{\raisebox{-.9pt} {\small 7}}}}
\end{split}
}
\label{pred_prod_form_13}
\end{equation}
Here,
by the indexed form for graph grammar rules of Equation~(\ref{equiv_rule_form}),
\begin{equation}
\begin{split}
\big[ \ldots \big]_{\raisebox{.5pt}{\textcircled{\raisebox{-.9pt} {\small 3}}}} 
	\big[ \ldots \big]_{\raisebox{.5pt}{\textcircled{\raisebox{-.9pt} {\small 4}}}} &= \hat{a}(G^{r_{1;2} \; \text{out}}) \\
\big[ \ldots \big]_{\raisebox{.5pt}{\textcircled{\raisebox{-.9pt} {\small 5}}}} 
	\big[ \ldots \big]_{\raisebox{.5pt}{\textcircled{\raisebox{-.9pt} {\small 6}}}} &= {a}(G^{r_{1;2} \; \text{in}}) \\
\end{split}
\end{equation}
as in the non-indexed semantics form of Equation~(\ref{eq1}).
Furthermore, the idempotent factors of $Y$ and $Z$
(diagonal in the number basis, multiplying each pure graph state by 0 or 1 )
just require sufficient free memory to operate
the ``churn'' of memory used in rule 1 and released in rule 2;
assuming index allocation works as designed from a countably infinite store,
it is equivalent in the sense of Equation~(\ref{particle-equivalence}) to drop these factors.
The Kronecker delta functions are interpreted as label-matching conditions 
in labelled graph matching as in Lemma 1, constraining the 1-1 correspondence map $h$
to respect the node labels and thus (again by Lemma 1)
to be an isomorphism of labelled graphs;
and they also help to ensure
that the normalization for number of equivalent outcome graphs is correct.
Thus in order to compute the labelled, numbered graph rewrite rule in each summand
over $S$ and $h$,
one needs to find a labelled subgraph $H$ of $G^{r_1 \, \text{out} }$
that is isomorphic as a labelled graph to 
a labelled subgraph $\tilde{H}$ of $G^{r_2 \, \text{in} }$,
and do this in an edge-maximal way;
then one needs to pick an isomorphism $h$ between $H$
and $\tilde{H}$; then using $H, \tilde{H}$ and $h$ to map between
the rule $r_1$ and $r_2$ node numberings,
one needs to compute the left hand side and right hand labelled graphs 
as numbered and labelled node sets and link sets.


Thus by careful interpretation of terms we arrive at the main result, except
limited to the case in which hanging edges are {\it not} removed by the rule semantics:
for the hanging-edge permissive semantics of Equations~(\ref{eq1}) and (\ref{eq2}),
or equivalently Equation~(\ref{equiv_rule_form}),
\begin{equation}
{\hat{W}}_{G^{r_2 \; \text{in}}  {\rightarrow} G^{r_2 \; \text{out}}  }
    {\hat{W}}_{G^{r_1 \; \text{in}}  {\rightarrow} G^{r_1 \; \text{out}}  }
	 \; \simeq \;
   \sum_{
   	\vbox{ 
		\hbox{\scriptsize $H \subseteq G^{r_1 \, \text{out} } \simeq \tilde{H} \subseteq G^{r_2 \, \text{in} }$ }
		 \hbox{\scriptsize \quad \;\;  | edge-maximal}
	}
     }
	\sum_{h: H \overset{\text{\tiny 1--1}}{  \hookrightarrow } \tilde{H}}
    \quad
{\hat{W}}_{G^{r_1 \; \text{in}} \cup (G^{r_2 \; \text{in}} \setminus \tilde{H}) 
	\underset{h}{\rightarrow}
	G^{r_2 \; \text{out}}  \cup (G^{r_1 \; \text{out}} \setminus H) } 
\label{main_peek}
\end{equation}

In more detail,
the summand graph rewrite rule is then defined by the disjoint unions $\dot\cup$
(reflecting time-reversal $ L \leftrightarrow R$ duality):
\begin{equation}
\boxed{
\begin{split}
G^{1;2 \; \text{in}}_{\text{nodes}}(\tilde{H}_{\text{nodes}}) &= G^{r_1 \; \text{in}}_{\text{nodes}} \dot\cup (G^{r_2 \; \text{in}}_{\text{nodes}} \setminus \tilde{H}_{\text{nodes}}) &
G^{1;2 \; \text{out}}_{\text{nodes}}(H_{\text{nodes}}) & = G^{r_2 \; \text{out}}_{\text{nodes}} \dot\cup (G^{r_1 \; \text{out}}_{\text{nodes}} \setminus H_{\text{nodes}}) \\
 & \equiv G^{r_1 \; \text{in}}_{\text{nodes}} \cup h^{-1 \star}(G^{r_2 \; \text{in}}_{\text{nodes}} \setminus \tilde{H}_{\text{nodes}}) &
  &\equiv G^{r_2 \; \text{out}}_{\text{nodes}} \cup h^{\star}(G^{r_1 \; \text{out}}_{\text{nodes}} \setminus H_{\text{nodes}}) \\
G^{1;2 \; \text{in}}_{\text{links}}(\tilde{H}_{\text{nodes}}) &= G^{r_1 \; \text{in}}_{\text{links}} \cup h^{-1 \star}(G^{r_2 \; \text{in}}_{\text{links}} \setminus \tilde{H}_{\text{links}}) &
G^{1;2 \; \text{out}}_{\text{links}}(H_{\text{nodes}})  &= G^{r_2 \; \text{out}}_{\text{links}} \cup h^{\star}(G^{r_1 \; \text{out}}_{\text{links}} \setminus H_{\text{links}}) \\
\end{split}
}
\label{compound_labelled_graph_defs_1}
\end{equation}
where 
$\dot\cup$ denotes disjoint union, and where
$h^{\star}$ extends $h$ by remapping the nodes of $G^{r_1}$
along $h$ if possible, and to the disjoint union nodes if not;
and likewise for $h^{-1}$.
%
The conserved core graphs are determined by shared node labels on the left and right of a rule:
\begin{equation}
\boxed{
\begin{split}
	K_a & = G^{r_a \; \text{in}}_{\text{nodes}}  \cap G^{r_a \; \text{out}}_{\text{nodes}} \\
	K_{1;2} &= 
	(K_1 \setminus H_{\text{nodes}}) 
	\; \cup \;  h^{-1}(K_2 \setminus  \tilde{H}_{\text{nodes}}) 
	 \; \cup \; (K_1 \cap h^{-1 \star}(K_2)) \\
\end{split}
}
\label{compound_labelled_graph_defs_2}
\end{equation}
The exact mechanics of graph numbering and disjoint union
are discussed in [1], and examples are given in 
Section~\ref{examples}.

Given the definitions of the compound label graphs 
in Equations~(\ref{compound_labelled_graph_defs_1}) and (\ref{compound_labelled_graph_defs_2}),
one can write the graph rewrite rule algebra as announced in Section \ref{advert_main_result}:

\begin{theorem}
For the hanging-edge-permissive semantics of Equations~(\ref{eq1}) and (\ref{eq2}),
or equivalently Equation~(\ref{equiv_rule_form}),
and assuming multiplicative normalization $C_r$, then
\begin{equation}
\boxed{
{\hat{W}}_{G^{r_2 \; \text{in}}  {\rightarrow} G^{r_2 \; \text{out}}  }
    {\hat{W}}_{G^{r_1 \; \text{in}}  {\rightarrow} G^{r_1 \; \text{out}}  }
	 \; \simeq \;
   \sum_{
   	\vbox{ 
		\hbox{\scriptsize $H \subseteq G^{r_1 \, \text{out} } \simeq \tilde{H} \subseteq G^{r_2 \, \text{in} }$ }
		 \hbox{\scriptsize \quad \;\;  | edge-maximal}
	}
     }
	\sum_{h: H \overset{\text{\tiny 1--1}}{  \hookrightarrow } \tilde{H}}
    \quad
{\hat{W}}_{G^{1;2 \; \text{in}}(\tilde{H}) 
	\underset{h}{\rightarrow}
	G^{1;2 \; \text{out}} (H) } 
}
\label{main_result}
\end{equation}
where the compound labelled graphs $G^{1;2 \; \text{in}}(\tilde{H})$ and $G^{1;2 \; \text{out}} (H)$,
and their label overlaps $K_{1;2}$, are defined by
Equations~(\ref{compound_labelled_graph_defs_1}) and (\ref{compound_labelled_graph_defs_2}) above.
The coefficients in this expression are all nonnegative integers
(as the same graph grammar rule
could arise several times by different means).
Rate factors $\rho$ multiply, as in Equation~(\ref{pred_prod_form_13}).
\end{theorem}


\begin{corollary}
There is an algebraic reduction of operator products to sums,
similar to Theorem 1,
that applies to the $W_r$ operators that subtract off
diagonal operators from $\hat{W}_r$ to conserve probability as in Equation~(\ref{MasterEqn}),
except that the coefficients can be any integer.
\end{corollary}

Proof:
$W_{G^{r_1 \; \text{in}} _\chi {\rightarrow} G^{r_1 \; \text{out}} _\chi} 
    = \hat{W}_{G^{r_1 \; \text{in}} _\chi {\rightarrow} G^{r_1 \; \text{out}} _\chi} 
        - \hat{W}_{G^{r_1 \; \text{in}} _\chi {\rightarrow} G^{r_1 \; \text{in}} _\chi}$ 
        (as shown in [1]) for $\chi \in \{r_1,r_2\}$;
the latter two terms are each subject to Theorem 1.
Thus $W_{r_2} W_{r_2} $ is equivalent to a sum of $\hat{W}_s$ operators for a set of labelled graph grammar rules indexed by $s$.
Since  $W_{r_2} $ preserves probability,
${\mathbf 1} \cdot W_{r_2} W_{r_1} = {\mathbf 0} \cdot W_{r_1}  = {\mathbf 0}$ .
We can therefore subtract zero in the form of $\text{diag}( {\mathbf 1} \cdot W_{r_2} W_{r_1})$,
applied term by term with the same sum of graph grammar rules substituted in for $W_{r_2} W_{r_1})$, 
and find that $W_{r_2} W_{r_2} $ is equivalent to a sum of $W=\hat{W}_s - \text{diag}( {\mathbf 1} \cdot  \hat{W}_s )$ 
operators for a set of labelled graph grammar rules indexed by $s$.

\begin{corollary}
There is an algebraic reduction of {\rm commutators} of labelled graph grammar rule state-change operators 
$\hat{W}_r$ to sums of the same form, 
similar to Theorem 1,
with integer coefficients. 
Also there is a similar algebraic reduction of commutators of  labelled graph grammar rule full operator
${W}_r$ commutators to sums of the same form, with integer coefficients.
\end{corollary}

Proof: As in Corollary 1, but with extra minus signs on some of the rule operators.

\begin{corollary}
There exists (as exhibited in the proof of Theorem 1) 
a {\rm constructive mapping} from the graph rewrite rule operator algebra semantics
to the elementary bitwise operator algebras of Section~\ref{bitwise_op_alg}.
Since it depends on an index allocation scheme which can be done in many ways,
this mapping is not unique.
\end{corollary}

\begin{corollary}
One particular subgraph that always contributes to the product is $H=\varnothing = \tilde{H}$, the empty graph.
Its contribution always cancels out of the commutator
$[{\hat{W}}_{r_2}, {\hat{W}}_{r_1} ] = {\hat{W}}_{r_2} {\hat{W}}_{r_1} - {\hat{W}}_{r_1} {\hat{W}}_{r_2}  $,
because 
then $H = \varnothing$ then
nothing is shared between the two rule firings so their order doesn't matter.
\end{corollary}

\begin{corollary}
Integrating  Equation~(\ref{main_result}) over parameters $\int d \mu_r(X) \ldots $,
as in Equation~(\ref{lambda_integral}),
results in a version of Equation~(\ref{main_result}) that incorporates parameter
integrals term-by-term.
\end{corollary}

Proof:
Using the delta functions in Equation~(\ref{pred_prod_form_13}),
$\int d \mu_{r_1} (X_1) d \mu_{r_2} (X_2) \prod \delta_{ \lambda^{\prime} \lambda} \ldots  $
integrates out some of the variables in $(X_1,X_2)$ via the delta functions
(Kronecker or Dirac depending on the measures $\mu_r$) 
but leaves behind others that assume the same form
$\int d \mu_r(X) \ldots $.

\subsection{Commutation calculation - with edge cleanup}
\label{calculation-cleanup}

We now turn to the hanging-edge cleanup semantics,
and prove (Theorem 2)
that the same algebra as in Theorem 1 and
 Equations~(\ref{compound_labelled_graph_defs_1}), (\ref{compound_labelled_graph_defs_2}), and (\ref{main_result}), 
 above still applies.
 
 The semantics is now
\begin{equation}
\begin{split}
{\hat{W}}_{r_\chi} \; =
{1 \over C_{r_\chi}(N_{\text{max free}})}
& \int d \mu_{r_\chi} X
\; \rho _{r_\chi}({\lambdab}[X] , {\lambdab^{\prime }}[X]) 
\sum \limits_{{\cal I}_\chi  : L_\chi \cup R_\chi \overset{\text{1-1}}{\hookrightarrow} {\cal U}}
\Bigg[ 
 \Big( 
	\prod \limits_{(i^{\prime}, \, i) \in {\cal P}_\chi }
		E_{i^{\prime} \; i}  \Big) 
  \Big(  
	\prod \limits_{(\hat{i}, \, \hat{i}^{\prime}) \in {\cal P}^*_\chi }
		E_{ \hat{i} \; \hat{i}^{\prime} } \Big) 
    \Bigg]  
\\ & \times
  \Bigg [ \prod \limits_{(i_1, i_2) \in {\cal R}_\chi } 
  	 {\hat{a}}_{i_1 i_2}
  \Bigg]  
  \Bigg [ \prod \limits_{(i_3, i_4) \in {\cal L}_\chi } 
  	 { {a}}_{i_3 i_4}
  \Bigg]  
  \Bigg [ \prod \limits_{i_5 \in  R_\chi }
  	  {\hat{a}}_{i_5 \, , \,  \lambda^{\prime}_{{\cal I}_\chi^{-1}(i_5)}}
  \Bigg]  
  \Bigg [ \prod \limits_{i_6 \in  L_\chi }
  	  {{a}}_{i_6 \, , \,  \lambda_{{\cal I}_\chi^{-1}(i_6)}}
  \Bigg]  .
\end{split}
\label{hanging_indexed_semantics}
\end{equation}

The product of two such operators is
(omitting for now the integral over parameters $X$)
\begin{equation}
\begin{split}
{\hat{W}}_{r_2} 
{\hat{W}}_{r_1} \; =
{1 \over C_{r_1}(N_{\text{max free}})}
  &
  {1 \over C_{r_2}(N_{\text{max free}})}
\rho _{r_1}({\lambdab}^{(1)}, {\lambdab^{\prime }}^{(1)}) 
\rho _{r_2}({\lambdab}^{(2)}, {\lambdab^{\prime }}^{(2)}) 
\\ & \times
\sum \limits_{{\cal J}  : L_2 \cup R_2 \overset{\text{1-1}}{\hookrightarrow} {\cal U}}
\quad 
\sum \limits_{{\cal I}  : L_1 \cup R_1 \overset{\text{1-1}}{\hookrightarrow} {\cal U}}
\Bigg[ 
 \Big( 
	\prod \limits_{(j^{\prime}, \, j) \in {\cal P}_2 }
		E_{j^{\prime} \; j}  \Big) 
  \Big(  
	\prod \limits_{(\hat{j}, \, \hat{j}^{\prime}) \in {\cal P}^*_2 }
		E_{ \hat{j} \; \hat{j}^{\prime} } \Big) 
    \Bigg]  
\\ & \times
  \Bigg [ \prod \limits_{(j_1, j_2) \in  {\cal R}_2 } 
  	 {\hat{a}}_{j_1 j_2}
  \Bigg]  
  \Bigg [ \prod \limits_{(j_3, j_4) \in  {\cal L}_2} 
  	 { {a}}_{j_3 j_4}
  \Bigg]  
  \Bigg [ \prod \limits_{j_5 \in  R_2 }
  	  {\hat{a}}_{j_5 \, , \,  \lambda^{\prime {(2)}}_{{\cal J}^{-1}(j_5)}}
  \Bigg]  
  \Bigg [ \prod \limits_{j_6 \in  L_2 }
  	  {{a}}_{j_6 \, , \,  \lambda^{(2)}_{{\cal J}^{-1}(j_6)}}
  \Bigg]  
\\ & \times
\Bigg[ 
 \Big( 
	\prod \limits_{(i^{\prime}, \, i) \in {\cal P}_1 }
		E_{i^{\prime} \; i}  \Big) 
  \Big(  
	\prod \limits_{(\hat{i}, \, \hat{i}^{\prime}) \in {\cal P}^*_1 }
		E_{ \hat{i} \; \hat{i}^{\prime} } \Big) 
    \Bigg]  
\\ & \times
  \Bigg [ \prod \limits_{(i_1, i_2) \in  {\cal R}_1} 
  	 {\hat{a}}_{i_1 i_2}
  \Bigg]  
  \Bigg [ \prod \limits_{(i_3, i_4) \in  {\cal L}_1} 
  	 { {a}}_{i_3 i_4}
  \Bigg]  
  \Bigg [ \prod \limits_{i_5 \in  R_1 }
  	  {\hat{a}}_{i_5 \, , \,  \lambda^{\prime {(1)}}_{{\cal I}^{-1}(i_5)}}
  \Bigg]  
  \Bigg [ \prod \limits_{i_6 \in  L_1 }
  	  {{a}}_{i_6 \, , \,  \lambda^{(1)}_{{\cal I}^{-1}(i_6)}}
  \Bigg]  
\end{split}
\label{pred_prod_form_1a}
\end{equation}
The problem is to treat the potentially very high degree factors of $\prod_{{\cal P}_1 \cup {\cal P}_1^*}  E$
 that have been inserted into the middle of this semantics.

\subsubsection{ Edge cleanup asymptotics}

We now work to replace the product of $E_{i j}$ factors above with the exponential of a sum.

\begin{equation}
\begin{split}
	E_\alpha = Z_\alpha + a_\alpha = I_\alpha + (a_\alpha - N_\alpha) = I_\alpha + W_{\alpha \rightarrow \varnothing} \\
\end{split}
\end{equation}
First we note an application of the Euler formula for the matrix exponential.
Defining
\begin{equation}
\tau = \rho_{\text{erase}} t  \; ,
\end{equation}
where $ \rho_{\text{erase}}$ is an effective high speed of interpolated edge erasures, then
\begin{equation}
\begin{split}
  \exp(\tau \sum_{\alpha \in {\cal S}}  W_{\alpha \rightarrow \varnothing} ) 
		&= 
		\lim_{m \rightarrow \infty} \Big( I + {\tau \over m}  \sum_{\alpha \in {\cal S}}  W_{\alpha \rightarrow \varnothing} \Big)^m \\
		&= 
		\lim_{m \rightarrow \infty} \Big( \prod_{\alpha \in {\cal S}} (I + {\tau \over m}    W_{\alpha \rightarrow \varnothing} ) \Big)^m \\
		&= 
		\prod_{\alpha \in {\cal S}}  \Big( \lim_{m \rightarrow \infty} (I + {\tau \over m}    W_{\alpha \rightarrow \varnothing} ) ^m\Big)
\end{split}
\end{equation}
where the product orders are arbitrary because different $W_{\alpha \rightarrow \varnothing}$ commute.
Defining $\epsilon = \tau/m$, another expression for this is
\begin{equation}
\begin{split}
  \exp(\tau  \sum_{\alpha \in {\cal S}}  W_{\alpha \rightarrow \varnothing} ) 
		&= 
		\lim_{m \rightarrow +\infty, \epsilon -> 0^+} \Big( \prod_{\alpha \in {\cal S}} (I +\epsilon    W_{\alpha \rightarrow \varnothing} ) \Big)^m .
\end{split}
\end{equation}

On the other hand,
recalling that $E_\alpha$, $N_\alpha$, $Z_\alpha$ and $I_\alpha$ are all idempotent ($E^2 = E$, etc.) in the 2 $\times$ 2 case,
\begin{equation}
\begin{split}
	\exp(\tau  \sum_{\alpha \in {\cal S}}  W_{\alpha \rightarrow \varnothing} ) 
	&=
	 \exp(\tau  \sum_{\alpha \in {\cal S}}  (E_\alpha - I_{\alpha}) ) \\
	 &=
	 \exp( -\tau  \sum_{\alpha \in {\cal S}}  I_{\alpha} )\exp(\tau  \sum_{\alpha \in {\cal S}}  E_{\alpha} ) \\
	 &=
	 \exp( -\tau  | {\cal S}| ) [ I +  \tau  \sum_{\alpha \in {\cal S}}  E_{\alpha} + 
	 	{ \tau^2 \over 2}[ \sum_{\alpha \in {\cal S}}  E_{\alpha}  +  \sum_{\alpha \neq \beta \in {\cal S}}  E_{\alpha} E_{\beta} ] + \cdots ]\\
	 &=
	 \exp( -\tau  | {\cal S}| ) 
	 	\Bigg[ \sum_{k=0}^\infty {\tau^k \over k!} 
	 	 \sum_{\alpha_1 \ldots \alpha_k \in {\cal S}}  E_{\alpha_1} \ldots E_{\alpha_k} \Bigg] \\
\end{split}
\end{equation}

Using $E_\alpha^2 = E_\alpha$ and grouping $\alpha_*$s into partition blocks of equal $\alpha$ value,
\begin{equation}
\begin{split}
	\exp(\tau  \sum_{\alpha \in {\cal S}}  W_{\alpha \rightarrow \varnothing} ) 
	 &=
	 \exp( -\tau  | {\cal S}| ) 
	 	\Bigg[ \sum_{k=0}^\infty {\tau^k \over k!} 
	 	 \sum_{\alpha_1 \ldots \alpha_k \in {\cal S}}  E_{\alpha_1} \ldots E_{\alpha_k} \Bigg] \\
	 &=
	 \exp( -\tau  | {\cal S}| ) 
	 	\Bigg[I + \sum_{k=1}^\infty {\tau^k \over k!} \;\;
			\sum_{l=1}^{\min(k,|{\cal S}| )} 
				 \stirlingii{k}{l}
	 	 \sum_{\langle \beta_1 \ldots \beta_l \in {\cal S} \rangle_{\neq}}  E_{\beta_1} \ldots E_{\beta_k} \Bigg] \\
	 &=
	 \exp( -\tau  | {\cal S}| ) 
	 	\Bigg[ I + 
			\sum_{l=1}^{|{\cal S} | }  \;\; 
				\Big[ \sum_{k=l}^\infty {\tau^k \over k!}
				 \stirlingii{k}{l}
	 	 \sum_{\langle \beta_1 \ldots \beta_l \in {\cal S} \rangle_{\neq}}  E_{\beta_1} \ldots E_{\beta_k} \Bigg] \\
	 &=
	 \exp( -\tau  | {\cal S}| ) 
	 	\Bigg[ I + 
			\sum_{l=1}^{|{\cal S} | }  \;\; 
				{ (e^{\tau } -1)^l \over l! }
	 	 \sum_{\langle \beta_1 \ldots \beta_l \in {\cal S} \rangle_{\neq}}  E_{\beta_1} \ldots E_{\beta_k} \Bigg] \\
\end{split}
\end{equation}
where  $ \stirlingii{k}{l}$
are Stirling numbers of the second kind and 
where the last line uses a generating function for these numbers.

Then asyptotically as $\tau = \rho_{\text{erase}} t \rightarrow +\infty$, and defining $|{\cal S}|_{(m)} \equiv |{\cal S}| ! / (|{\cal S}| -m)!$,
where $m+l = |{\cal S}|$,
\begin{equation}
\begin{split}
	\exp(\tau  \sum_{\alpha \in {\cal S}}  W_{\alpha \rightarrow \varnothing} ) 
 & \rightarrow
 	 \exp( -\tau  | {\cal S}| ) 
	 	 I + {1 \over |{\cal S}|!}
			\sum_{m=0}^{|{\cal S} | -1}  \;\; 
				|{\cal S}|_{(m)} e^{-m t}
	 	 \sum_{\langle \beta_1 \ldots \beta_{|{\cal S}|-m} \in {\cal S} \rangle_{\neq}}  E_{\beta_1} \ldots E_{\beta_{|{\cal S}|-m}}  \\
 & \rightarrow
 	 {1 \over |{\cal S}|!}
	 	 \sum_{\langle \beta_1 \ldots \beta_{|{\cal S}|} \in {\cal S} \rangle_{\neq}}  E_{\beta_1} \ldots E_{\beta_{|{\cal S}|}}  \\
 &=
 	\prod_{\alpha \in {\cal S} } E_{\alpha} .
\end{split}
\end{equation}
So, {\it complete erasure is the limiting behavior of this edge-by-edge stochastic erasure process},
and it can be achieved simply by taking the limit $\rho_{\text{erase}} \rightarrow +\infty$.

Now we apply these calculations to the actual hanging-edge erasure operator:
\begin{equation}
\begin{split}
	\exp(\tau  \sum_{(i_1, i_2) \in {\cal S}}  W_{(i_1, i_2)  \rightarrow \varnothing} ) 
 & =
	\exp (\tau  \sum_{(i_1, i_2) \in {\cal S}}  ( E_{ i_1, i_2} - I_{i_1, i_2} ) N_{ i_2}  Z_{i_1}   )  \\
\end{split}
\label{exp_erasure_reduction}
\end{equation}
Here the node operator $Z_i $ checks for unallocated nodes $i$ with no label:
\begin{equation}
\begin{split}
Z_i \equiv& \prod_\lambda Y_{i \, , \lambda} = N_{i \, , \varnothing} \prod_\lambda Z_{i \, , \lambda} \\
&= N_{i \, , \varnothing} \prod_\lambda (I- N_{i \, , \lambda}) \\
&\simeq N_{i \, , \varnothing} (I - \sum_\lambda  N_{i \, , \lambda} )
    & \text{(since $N_{i \, , *}$ cross-terms vanish from WTA)} \\
&\simeq N_{i \, , \varnothing} \cdot N_{i \, , \varnothing} 
    & \text{(from Equation~(\ref{eq7_3_2}), top line)}  \\
    Z_i  &= N_{i \, , \varnothing} \\
\end{split}
\label{erasure_reduction}
\end{equation}
whence $Z_i Z_i = Z_i$.
Also $N_i \equiv \sum_\lambda N_{i \, , \lambda}$ counts the number of active labels for node $i$
which by WTA constraint is $0$ or $1$; we have again $N_i N_i = N_i$ and $N_i+Z_i = I$ and $N_i Z_i = 0$.
We note here that the operator $Z$ in Equation~(\ref{erasure_reduction}) {\it doesn't} quite fit within
the graph grammar rule semantics we have defined so far because it checks for {\it non}existence.
Nonexistence checks are 
identified as a more general kind of semantics in  [5] and [1],
which we do not treat in the present work except for this particular technical example.
Of course, Equation~(\ref{exp_erasure_reduction})  doesn't need to fit within the rule semantics, 
as it is not
explicitly accessible at the level of stochastic labelled graph grammar rules - it is just substructure.

Again
defining $\epsilon = \tau /m$, another expression for the exponential in Equation~(\ref{erasure_reduction})  is
\begin{equation}
\begin{split}
  \exp(\tau  \sum_{\alpha \in {\cal S}}  W_{ (i_1, i_2) \in {\cal S} \rightarrow \varnothing} ) 
		&= 
		\lim_{m \rightarrow +\infty, \epsilon -> 0^+} \Big( \prod_{(i_1, i_2) \in {\cal S}} (I +\epsilon    W_{(i_1, i_2) \in {\cal S} \rightarrow \varnothing} ) \Big)^m \\
\end{split}
\end{equation}

On the other hand,
\begin{equation}
\begin{split}
	\exp(\tau  \sum_{(i, j) \in {\cal S}}  W_{(i_1, i_2)  \rightarrow \varnothing} ) 
 & =
	\exp (\tau  \sum_{(i, j) \in {\cal S}}  ( E_{i, j} - I_{i, j} ) N_{ j}  Z_{i}   )  \\
	 &=
	 \exp( -\tau  \sum_{(i, j) \in {\cal S}}  N_{ j}  Z_{i} )\exp(\tau  \sum_{(i_1, i_2) \in {\cal S}}  E_{i, j} N_{ j}  Z_{i}) \\
	 &=
	 \exp( -\tau  | {\cal S}| ) 
	 	\Bigg[ \sum_{k=0}^\infty {\tau^k \over k!} 
	 	 \sum_{(i_1, j_1) \ldots (i_k, j_k) \in {\cal S}}  (E_{i_1, j_1} \ldots E_{i_k, j_k} ) (N_{ j_1} \ldots N_{ j_k} ) (Z_{i_1} \ldots Z_{i_k} ) \Bigg] \\
	 &=
	 \exp( -\tau  | {\cal S}| ) 
	 	\Bigg[I + \sum_{k=1}^\infty {\tau^k \over k!} \;\;
			\sum_{l=1}^{\min(k,|{\cal S}| )} 
				 \stirlingii{k}{l}
				 \\ & \quad \quad
	 	 \sum_{\langle (i_1, j_1) \ldots (i_l, j_l) \in {\cal S}  \rangle_{\neq}}   (E_{i_1, j_1} \ldots E_{i_l, j_l} ) (N_{ j_1} \ldots N_{ j_l} ) (Z_{i_1} \ldots Z_{i_l} ) \Bigg] \\
	 &=
	 \exp( -\tau  | {\cal S}| ) 
	 	\Bigg[ I + 
			\sum_{l=1}^{|{\cal S} | }  \;\; 
				\Big[ \sum_{k=l}^\infty {\tau^k \over k!}
				 \stirlingii{k}{l}
				 \\ & \quad \quad
	 	 \sum_{\langle (i_1, j_1) \ldots (i_l, j_l) \in {\cal S}  \rangle_{\neq}}   (E_{i_1, j_1} \ldots E_{i_l, j_l} ) (N_{ j_1} \ldots N_{ j_l} ) (Z_{i_1} \ldots Z_{i_l} ) \Bigg] \\
	 &=
	 \exp( -\tau  | {\cal S}| ) 
	 	\Bigg[ I + 
			\sum_{l=1}^{|{\cal S} | }  \;\; 
				{ (e^\tau -1)^l \over l! }
				 \\ & \quad \quad
	 	 \sum_{\langle (i_1, j_1) \ldots (i_l, j_l) \in {\cal S}  \rangle_{\neq}}   (E_{i_1, j_1} \ldots E_{i_l, j_l} ) (N_{ j_1} \ldots N_{ j_l} ) (Z_{i_1} \ldots Z_{i_l} ) \Bigg] \\
\end{split}
\end{equation}
where  as before 
$ \stirlingii{k}{l}$
 are Stirling numbers of the second kind and 
where the last line uses a generating function for these numbers.
Then asyptotically as $\tau = \rho_{\text{erase}} t \rightarrow +\infty$, and defining $|{\cal S}|_{(m)} \equiv |{\cal S}| ! / (|{\cal S}| -m)!$,
\begin{equation}
\begin{split}
	\exp(\tau  \sum_{\alpha \in {\cal S}}  W_{\alpha \rightarrow \varnothing} ) 
 & \rightarrow
 	 \exp( -\tau  | {\cal S}| ) 
	 	 I + {1 \over |{\cal S}|!}
			\sum_{m=0}^{|{\cal S} | -1}  \;\; 
				|{\cal S}|_{(m)} e^{-m\tau }
				 \\ & \quad \quad
	 	 \sum_{\langle (i_1, j_1) \ldots (i_l, j_l) \in {\cal S}  \rangle_{\neq}}   (E_{i_1, j_1} \ldots E_{i_l, j_l} ) (N_{ j_1} \ldots N_{ j_l} ) (Z_{i_1} \ldots Z_{i_l} ) \\
 & \rightarrow
 	 {1 \over |{\cal S}|!}
	 	 \sum_{\langle (i_1, j_1) \ldots (i_l, j_l) \in {\cal S}  \rangle_{\neq}}   (E_{i_1, j_1} \ldots E_{i_l, j_l} ) (N_{ j_1} \ldots N_{ j_l} ) (Z_{i_1} \ldots Z_{i_l} ) \\
 &=
 	\prod_{(i,j) \in {\cal S} } E_{i, j} N_j Z_i \\
	& \simeq
 	\prod_{(i,j) \in {\cal P} } E_{i, j} \; .
\end{split}
\end{equation}
The final line above is a key step prepared for by the discussion in
Section~\ref{hanging_edge_cleanup},
and it is justified by the fact that inductively the operator $N_j$ 
produces a zero value unless node $j$ has been allocated
at some point in the history of rule-firings .

So again we get the product of forward edge erasures by an incremental process of deletion, run for a long effective time $\tau$.

\subsubsection{Commutation with edge cleanup}

In Equation~(\ref{pred_prod_form_1a}), as in (\ref{Edge cleanup}),
\begin{equation}
\begin{split}
\hat{W}^{\text{cleaned}} 
&= \Big( \prod_{(k_1,k_2) \in {\cal S}} E_{k_1 k_2} E_{k_2 k_1} \Big) \hat{W}^{\text{bare}} \\
&= \lim_{n \rightarrow +\infty, \epsilon \rightarrow 0^+}
	[I + \epsilon \sum_{(k_1,k_2) \in {\cal S}} (a_{k_1,k_2} - N_{k_1,k_2}) N_{k_2} Z_{k_1}]^n
	[I + \epsilon \sum_{(k_1,k_2) \in {\cal S}} (a_{k_2,k_1} - N_{k_2,k_1}) N_{k_1} Z_{k_2}]^n
		\hat{W}^{\text{bare}} \\
\end{split}
\end{equation}
The core calculation within $\hat{W}_{r_2}^{\text{cleaned}} \cdot \hat{W}_{r_1}^{\text{cleaned}} $ is thus:
\begin{equation}
\begin{split}
	\hat{W}_{r_2}^{\text{bare}} 
	[\epsilon & \sum_{(k_1,k_2) \in {\cal S}} (a_{k_1,k_2} - N_{k_1,k_2}) N_{k_2} Z_{k_1}]
	\\ & =
	{\epsilon \over C_{r_2}} \sum_{\cal I} \sum_{(k_1,k_2) \in {\cal S}} 
		\Big[\prod_{{(i_1, i_2)} \in {\cal R}_2} \hat{a}_{i_1 i_2} \Big]
		\Big[\prod_{{(i_3, i_4)} \in {\cal L}_2} {a}_{i_3 i_4} \Big]
		(a_{k_1,k_2} - N_{k_1,k_2})
		\\ & \quad \quad \times
		\Big[\prod_{{i_5} \in R_2} \hat{a}_{i_5, \lambda_{{\cal I}^{-1}(i_5)}} \Big]
		\Big[\prod_{{i_6} \in L_2} {a}_{i_6, \lambda_{{\cal I}^{-1}(i_6)}} \Big] N_{k_2} Z_{k_1}
\end{split}
\end{equation}
Now calculate components:

{\bf Nodes}:
\begin{equation}
\begin{split}
\Big[ \prod_{{i_6} \in L_2} {a}_{i_6, \lambda_{{\cal I}^{-1}(i_6)}} \Big] N_{k_2}
	&= 
	\begin{cases}
		 [ \prod_{{i_6} \in L_2} {a}_{i_6, \lambda_{{\cal I}^{-1}(i_6)}}] & \text{if } k_2 \in L_2 \\
		 N_{k_2}  [ \prod_{{i_6} \in L_2} {a}_{i_6, \lambda_{{\cal I}^{-1}(i_6)}}] & \text{if } k_2 \notin L_2 
	\end{cases}
\end{split}
\end{equation}
\begin{equation}
\begin{split}
\Big[ \prod_{{i_5} \in R_2} \hat{a}_{i_5, \lambda_{{\cal I}^{-1}(i_5)}} \Big] N_{k_2}
	&= 
	\begin{cases}
		0 & \text{if } k_2 \in { R}_2 \\
		 N_{k_2}  [\prod_{{i_5} \in R_2} \hat{a}_{i_5, \lambda_{{\cal I}^{-1}(i_5)}} ] & \text{if } k_2 \notin {R }_2 
	\end{cases}
\end{split}
\end{equation}
so
\begin{equation}
\begin{split}
\Big[ \prod_{{i_5} \in R_2} \hat{a}_{i_5, \lambda_{{\cal I}^{-1}(i_5)}} \Big] 
  \Big[ \prod_{{i_6} \in L_2} {a}_{i_6, \lambda_{{\cal I}^{-1}(i_6)}} \Big] 
     N_{k_2}
	&= 
	\begin{cases}
		0 & \text{if } k_2 \in { R}_2 \setminus L_2 \\
		 \Big[ \prod_{{i_5} \in R_2} \hat{a}_{i_5, \lambda_{{\cal I}^{-1}(i_5)}} \Big]  
		 	\Big[ \prod_{{i_6} \in L_2} {a}_{i_6, \lambda_{{\cal I}^{-1}(i_6)}} \Big] 
			& \text{if } k_2 \in L_2 \\
		 N_{k_2}  \Big[ \prod_{{i_5} \in R_2} \hat{a}_{i_5, \lambda_{{\cal I}^{-1}(i_5)}} \Big]   
		 	\Big[ \prod_{{i_6} \in L_2} {a}_{i_6, \lambda_{{\cal I}^{-1}(i_6)}} \Big] 
			& \text{if } k_2 \in \overline{L_2} \cap \overline{R_2}
	\end{cases}
\end{split}
\end{equation}
Likewise $Z_k = I - N_k \implies N_k = I - Z_k$ and
\begin{equation}
\begin{split}
\Big[ \prod_{{i_6} \in L_2} {a}_{i_6, \lambda_{{\cal I}^{-1}(i_6)}} \Big] Z_{k_1}
	&= 
	\begin{cases}
		 0 & \text{if } k_1 \in L_2 \\
		 Z_{k_1} [ \prod_{{i_6} \in L_2} {a}_{i_6, \lambda_{{\cal I}^{-1}(i_6)}}] & \text{if } k_2 \notin L_2 
	\end{cases}
\end{split}
\end{equation}
\begin{equation}
\begin{split}
\Big[ \prod_{{i_5} \in R_2} \hat{a}_{i_5, \lambda_{{\cal I}^{-1}(i_5)}} \Big] Z_{k_1}
	&= 
	\begin{cases}
		 [\prod_{{i_5} \in R_2} \hat{a}_{i_5, \lambda_{{\cal I}^{-1}(i_5)}} ] & \text{if } k_2 \in R_2 \\
		 Z_{k_1} [\prod_{{i_5} \in R_2} \hat{a}_{i_5, \lambda_{{\cal I}^{-1}(i_5)}} ] & \text{if } k_2 \notin R_2 
	\end{cases}
\end{split}
\end{equation}
so
\begin{equation}
\begin{split}
\Big[ \prod_{{i_5} \in R_2} \hat{a}_{i_5, \lambda_{{\cal I}^{-1}(i_5)}} \Big] 
  \Big[ \prod_{{i_6} \in L_2} {a}_{i_6, \lambda_{{\cal I}^{-1}(i_6)}} \Big] 
     Z_{k_1}
	&= 
	\begin{cases}
		0 & \text{if } k_1 \in L_2 \\
		 \Big[ \prod_{{i_5} \in R_2} \hat{a}_{i_5, \lambda_{{\cal I}^{-1}(i_5)}} \Big]  
		 	\Big[ \prod_{{i_6} \in L_2} {a}_{i_6, \lambda_{{\cal I}^{-1}(i_6)}} \Big] 
			& \text{if } k_1 \in R_2 \setminus L_2 \\
		 Z_{k_1}  \Big[ \prod_{{i_5} \in R_2} \hat{a}_{i_5, \lambda_{{\cal I}^{-1}(i_5)}} \Big]   
		 	\Big[ \prod_{{i_6} \in L_2} {a}_{i_6, \lambda_{{\cal I}^{-1}(i_6)}} \Big] 
			& \text{if } k_1 \in \overline{L_2} \cap \overline{R_2}
	\end{cases}
\end{split}
\end{equation}

Together, then, 
\begin{equation}
\boxed{
\begin{split}
\Big[ \prod_{{i_5} \in R_2} \hat{a}_{i_5, \lambda_{{\cal I}^{-1}(i_5)}} \Big] & 
  \Big[ \prod_{{i_6} \in L_2} {a}_{i_6, \lambda_{{\cal I}^{-1}(i_6)}} \Big] 
     N_{k_2} Z_{k_1} \\
	&= 
	\begin{cases}
		0 & \text{if } (k_1 \in L_2) \lor (k_2 \in R_2 \setminus L_2) \\
		 \Big[ \prod_{{i_5} \in R_2} \hat{a}_{i_5, \lambda_{{\cal I}^{-1}(i_5)}} \Big]  
		 	\Big[ \prod_{{i_6} \in L_2} {a}_{i_6, \lambda_{{\cal I}^{-1}(i_6)}} \Big] 
			& \text{if } (k_1 \in R_2 \setminus L_2) \land (k_2 \in L_2)  \\
		 Z_{k_1}  \Big[ \prod_{{i_5} \in R_2} \hat{a}_{i_5, \lambda_{{\cal I}^{-1}(i_5)}} \Big]   
		 	\Big[ \prod_{{i_6} \in L_2} {a}_{i_6, \lambda_{{\cal I}^{-1}(i_6)}} \Big] 
			& \text{if } (k_1 \in \overline{L_2} \cap \overline{R_2}) \land (k_2 \in L_2) \\
		 N_{k_2}   \Big[ \prod_{{i_5} \in R_2} \hat{a}_{i_5, \lambda_{{\cal I}^{-1}(i_5)}} \Big]   
		 	\Big[ \prod_{{i_6} \in L_2} {a}_{i_6, \lambda_{{\cal I}^{-1}(i_6)}} \Big] 
			& \text{if }  (k_1 \in R_2 \setminus L_2) \land  (k_2 \in \overline{L_2} \cap \overline{R_2}) \\
		 N_{k_2}  Z_{k_1}  \Big[ \prod_{{i_5} \in R_2} \hat{a}_{i_5, \lambda_{{\cal I}^{-1}(i_5)}} \Big]   
		 	\Big[ \prod_{{i_6} \in L_2} {a}_{i_6, \lambda_{{\cal I}^{-1}(i_6)}} \Big] 
			& \text{if } (k_1 \in \overline{L_2} \cap \overline{R_2}) \land (k_2 \in \overline{L_2} \cap \overline{R_2}) .
	\end{cases}
\end{split}
}
\label{together_nodes}
\end{equation}

{\bf Links}: We continue to calculate
\begin{equation}
\begin{split}
\Big[ \prod_{{(i_3, i_4)} \in {\cal L}_2} {a}_{i_3 i_4} \Big] a_{k_1,k_2}
	&= 
	\begin{cases}
		0 & \text{if } (k_1,k_2) \in {\cal L}_2 \\
		 a_{k_1,k_2}  [\prod_{{(i_3, i_4)} \in {\cal L}_2} {a}_{i_3 i_4} ] & \text{if } (k_1,k_2) \notin {\cal L}_2 
	\end{cases}
\end{split}
\label{deepcalc1}
\end{equation}
and	
\begin{equation}
\begin{split}
\Big[  \prod_{{(i_1, i_2)} \in {\cal R}_2} \hat{a}_{i_1 i_2}   \Big] a_{k_1,k_2}
	&= 
	\begin{cases}
		 N_{k_1,k_2}  [ \prod_{{(i_1, i_2)} \in {\cal R}_2 \setminus (i_1, i_2)} \hat{a}_{i_1 i_2}  ]& \text{if } (k_1,k_2) \in {\cal R}_2 \\
		 a_{k_1,k_2}  [  \prod_{{(i_1, i_2)} \in {\cal R}_2} \hat{a}_{i_1 i_2}  ] &\text{if } (k_1,k_2) \notin {\cal R}_2 
	\end{cases}
\end{split}
\label{deepcalc2}
\end{equation}

Next,
\begin{equation}
\begin{split}
\Big[ \prod_{{(i_3, i_4)} \in {\cal L}_2} {a}_{i_3 i_4} \Big] \hat{a}_{k_1,k_2} 
	&= 
	\begin{cases}
		Z_{k_1,k_2}  [\prod_{{(i_3, i_4)} \in {\cal L}_2 \setminus (k_1,k_2) } {a}_{i_3 i_4} ] & \text{if } (k_1,k_2) \in {\cal L}_2 \\
		\hat{a}_{k_1,k_2}  [\prod_{{(i_3, i_4)} \in {\cal L}_2} {a}_{i_3 i_4} ] & \text{if } (k_1,k_2) \notin {\cal L}_2 
	\end{cases}
\end{split}
\label{deepcalc3}
\end{equation}
and since $Z_\alpha a_\alpha = a_\alpha$,
\begin{equation}
\begin{split}
\Big[ \prod_{{(i_3, i_4)} \in {\cal L}_2} {a}_{i_3 i_4} \Big] N_{k_1,k_2} 
	&= 
	\Big[ \prod_{{(i_3, i_4)} \in {\cal L}_2} {a}_{i_3 i_4} \Big] \hat{a}_{k_1,k_2}  {a}_{k_1,k_2} &= 
	\begin{cases}
		 [\prod_{{(i_3, i_4)} \in {\cal L}_2} {a}_{i_3 i_4} ] & \text{if } (k_1,k_2) \in {\cal L}_2 \\
		N_{k_1,k_2}  [\prod_{{(i_3, i_4)} \in {\cal L}_2} {a}_{i_3 i_4} ] & \text{if } (k_1,k_2) \notin {\cal L}_2 
	\end{cases}
\end{split}
\label{deepcalc4}
\end{equation}
so from Equations~(\ref{deepcalc1}) and (\ref{deepcalc4}),
%
\begin{equation}
\begin{split}
\Big[ \prod_{{(i_3, i_4)} \in {\cal L}_2} {a}_{i_3 i_4} \Big] ({a}_{k_1,k_2}  - N_{k_1,k_2} )
	&= 
	\begin{cases}
		-  [\prod_{{(i_3, i_4)} \in {\cal L}_2} {a}_{i_3 i_4} ] & \text{if } (k_1,k_2) \in {\cal L}_2 \\
		(a_{k_1,k_2} - N_{k_1,k_2} ) [\prod_{{(i_3, i_4)} \in {\cal L}_2} {a}_{i_3 i_4} ] & \text{if } (k_1,k_2) \notin {\cal L}_2 
	\end{cases}
\end{split}
\label{deepcalc5}
\end{equation}

Likewise:
\begin{equation}
\begin{split}
\Big[  \prod_{{(i_1, i_2)} \in {\cal R}_2} \hat{a}_{i_1 i_2}   \Big] N_{k_1,k_2}
	&= 
	\begin{cases}
		 0 & \text{if } (k_1,k_2) \in {\cal R}_2 \\
		 N_{k_1,k_2}  [  \prod_{{(i_1, i_2)} \in {\cal R}_2} \hat{a}_{i_1 i_2}  ] &\text{if } (k_1,k_2) \notin {\cal R}_2 
	\end{cases}
\end{split}
\label{deepcalc6}
\end{equation}
so from Equations~(\ref{deepcalc2}) and (\ref{deepcalc5}),
\begin{equation}
\begin{split}
\Big[ 	\prod_{{(i_1, i_2)} \in {\cal R}_2} \hat{a}_{i_1 i_2}  \Big] ({a}_{k_1,k_2}  - N_{k_1,k_2} )
	&= 
	\begin{cases}
		N_{k_1,k_2}  [ \prod_{{(i_1, i_2)} \in {\cal R}_2} \hat{a}_{i_1 i_2}  ] & \text{if } (k_1,k_2) \in {\cal R}_2 \\
		(a_{k_1,k_2} - N_{k_1,k_2} ) [ \prod_{{(i_1, i_2)} \in {\cal R}_2} \hat{a}_{i_1 i_2}  ]  & \text{if } (k_1,k_2) \notin {\cal R}_2 
	\end{cases}
\end{split}
\end{equation}
Combining,
\begin{equation}
\boxed{
\begin{split}
\Big[ 	\prod_{{(i_1, i_2)} \in {\cal R}_2} \hat{a}_{i_1 i_2}  \Big] &
  \Big[ \prod_{{(i_3, i_4)} \in {\cal L}_2} {a}_{i_3 i_4} \Big]
      ({a}_{k_1,k_2}  - N_{k_1,k_2} ) \\
	&= 
	\begin{cases}
		-  \Big[ 	\prod_{{(i_1, i_2)} \in {\cal R}_2} \hat{a}_{i_1 i_2}  \Big]
		   \Big[ \prod_{{(i_3, i_4)} \in {\cal L}_2} {a}_{i_3 i_4} \Big]
			 & \text{if } (k_1,k_2) \in {\cal L}_2 \\
		N_{k_1,k_2}  \Big[ \prod_{{(i_1, i_2)} \in {\cal R}_2} \hat{a}_{i_1 i_2}  \Big]
		   \Big[ \prod_{{(i_3, i_4)} \in {\cal L}_2} {a}_{i_3 i_4} \Big]
			 & \text{if } (k_1,k_2) \in {\cal R}_2 \setminus {\cal L}_2 \\
		(a_{k_1,k_2} - N_{k_1,k_2} )  \Big[ \prod_{{(i_1, i_2)} \in {\cal R}_2} \hat{a}_{i_1 i_2}  \Big]
		    \Big[ \prod_{{(i_3, i_4)} \in {\cal L}_2} {a}_{i_3 i_4} \Big]
		     & \text{if } (k_1,k_2) \in \overline{\cal R}_2 \cap  \overline{\cal L}_2 .
	\end{cases}
\end{split}
}
\label{link_combined}
\end{equation}
	
Next we argue:
If
$(k_1,k_2) \in \overline{\cal R}_2 \cap  \overline{\cal L}_2$
as in the third line of the right hand side of Equation~(\ref{link_combined}) above
then the commutation was successful, and the factor of $a-N$ simply joins the infinite supply
of such factors to the left.
That leaves two cases in Equation~(\ref{link_combined}).
If $(k_1,k_2) \in {\cal L}_2$ 
(as in the first line of  the right hand side of Equation~(\ref{link_combined}) above)
then
$k_1 \in L_2 \wedge k_2 \in L_2$
so in Equation~(\ref{together_nodes})
the first line applies and the term is zero; it doesn't contribute.
That leaves one case:
$ (k_1,k_2) \in {\cal R}_2 \setminus {\cal L}_2 $,
in which case 
$k_1 \in R_2 \wedge k_2 \in R_2$.
Then either the first line  the right hand side of in Equation~(\ref{together_nodes}) again
applies and eliminates the present term, or else neither of its alternative conditions apply and
$k_1 \in R_2 \setminus L_2 \wedge k_2 \in R_2 \cap L_2$;
%
%
thus the condition for a surviving prefactor of $N_{k_1,k_2}$ is:
\begin{equation}
(k_1, k_2) \in [R_2 \setminus L_2 \times L_2 \cap R_2] \cap ({\cal R}_2 \setminus {\cal L}_2)
\end{equation}
... a condition excluded by the index allocation scheme, which implies $k_1 \notin R_2 \setminus L_2$.
So, all surviving terms behave as in the third line of Equation~(\ref{link_combined}),
and the factor of $a-N$ to the right of the second rule firing simply joins the infinite supply
of such factors to its left.

Intuitively, this means that hanging edges can be eliminated at any time rather
than promptly after every rule firing. This is because the assumed form of the graph
rewrite rules does not recognize or respond to hanging edges;
all edges are verified to have two vertices before a rule can fire.
As an aside, this explanation would not be valid if another alternative semantics
considered in [1] were used in which, like the nonconforming operator
$W_{(i_1,i_2) \rightarrow \varnothing}$ above, 
the LHS of a rule could test for nonexistence as well as existence of some entities.
Then a more complex algebraic operator equation might result.

Thus we find no change to the algebraic formula of Theorem 1 for the 
hanging-edge removal semantics:

\begin{theorem}
For the hanging-edges removal semantics of 
Equations~(\ref{Edge_cleanup}) and (\ref{Edge_cleanup_full}),
or equivalently Equation~(\ref{hanging_indexed_semantics}),
and assuming finiteness of rules, index allocation blocks,
and number of rule firings,
and assuming multiplicative normalization $C_r$, then
\begin{equation}
{\hat{W}}_{G^{r_2 \; \text{in}}  {\rightarrow} G^{r_2 \; \text{out}}  }
    {\hat{W}}_{G^{r_1 \; \text{in}}  {\rightarrow} G^{r_1 \; \text{out}}  }
	 \; \simeq \;
   \sum_{
   	\vbox{ 
		\hbox{\scriptsize $H \subseteq G^{r_1 \, \text{out} } \simeq \tilde{H} \subseteq G^{r_2 \, \text{in} }$ }
		 \hbox{\scriptsize \quad \;\;  | edge-maximal}
	}
     }
	\sum_{h: H \overset{\text{\tiny 1--1}}{  \hookrightarrow } \tilde{H}}
    \quad
{\hat{W}}_{G^{1;2 \; \text{in}}(\tilde{H}) 
	\underset{h}{\rightarrow}
	G^{1;2 \; \text{out}} (H) } 
\label{main_result_2}
\end{equation}
where the compound labelled graphs $G^{1;2 \; \text{in}}(\tilde{H})$ and $G^{1;2 \; \text{out}} (H)$,
and their label overlaps $K_{1;2}$, are defined by
Equations~(\ref{compound_labelled_graph_defs_1}) and (\ref{compound_labelled_graph_defs_2}) above.
The coefficients in this expression are all nonnegative integers
(as the same graph grammar rule
could arise several times by different means).
Rate factors $\rho$ multiply, as in Equation~(\ref{pred_prod_form_13}).
\end{theorem}

\begin{corollary}
There is an algebraic reduction of operator products to sums,
similar to Theorem 2,
that applies to the $W_r$ operators that subtract off
diagonal operators from $\hat{W}_r$ to conserve probability,
except that the coefficients can be any integer.
\end{corollary}

{\it Proof:} Exactly as for Corollary 1.

\begin{corollary}
There is an algebraic reduction of {\rm commutators} of labelled graph grammar rule state-change operators 
$\hat{W}_r$ to sums of the same form, 
similar to Theorem 2,
with integer coefficients. 
Also there is a similar algebraic reduction of commutators of  labelled graph grammar rule full operator
${W}_r$ commutators to sums of the same form, with integer coefficients.
\end{corollary}

Proof: As in Corollary 6 or 1, but with extra minus signs on some of the rule operators.

\begin{corollary}
There exists (as exhibited in the proofs of Theorem 1 and 2) 
a {\rm constructive mapping} from the graph rewrite rule operator algebra semantics
to the elementary bitwise operator algebras of Section~\ref{bitwise_op_alg}.
Since it depends on a index allocation scheme which can be done in many ways,
this mapping is not unique.
\end{corollary}

\begin{corollary}
One particular subgraph that always contributes to the product is $H=\varnothing = \tilde{H}$, the empty graph.
Its contribution always cancels out of the commutator
$[{\hat{W}}_{r_2}, {\hat{W}}_{r_1} ] = {\hat{W}}_{r_2} {\hat{W}}_{r_1} - {\hat{W}}_{r_1} {\hat{W}}_{r_2}  $,
because nothing is shared between the two rule firings so their order doesn't matter.
\end{corollary}

We note here that a previous attempt to prove Theorem 2 directly 
using the large product of $E$ operators and ${\cal P}, L, R, {\cal L, R}$ etc.
by Boolean logic ran aground in notational complexity.
The method used here, with the exponential of a sum of $E - I$ operators, seems more tractable.

\begin{corollary}
Integrating  Equation~(\ref{main_result_2}) over parameters $\int d \mu_r(X) \ldots $,
as in Equation~(\ref{lambda_integral}),
results in a version of Equation~(\ref{main_result}) that incorporates parameter
integrals term-by-term.
\end{corollary}

{\it Proof:} Exactly as for Corollary 5.

\subsection{Discussion of Theorems}

In Lie group theory, the Lie algebra is related to the curvature tensor of a group-invariant metric.
The Lie algebras discussed here are in much higher dimension
but may also relate to geometric or topological structures.

As discussed in [1], operator commutators provide an analytic tool,
when used with perturbation series expansions such as
the Baker-Campbell-Hausdorff theorem
(as suggested for rewrite operator algebras in [5, 13])
underlying operator splitting methods,
or the Time-Ordered Product Expansion for Feynman diagrams underlying the Gillespie Stochastic Simulation Algorithm
and some of its generalizations [10],
by which to derive both general and model-specific simulation simulation algorithms
and to estimate and/or bound their errors.

By way of comparison with other work,
there is an alternative category-theory based
approach to graph grammar semantics 
based on single or double pushout commutative diagrams rather than operator algebras,
and a collection of ``independence'' conditions
for two successive rule firings to have an order-independent result
[14]. 
In our operator algebra language these conditions would guarantee
a zero commutator.
The work of [15, 13] combines 
and connects together both double-pushout and 
master equation semantics, using a restricted subset of the operator algebra 
implied by Propositions 1 or 2 of [1] or the more powerful
Theorems 1 and 2 of this work.

\section{Examples and Discussion}
\label{examples}

A minority of biological models have been formulated in terms of structural rewrite rules
for graphs and cell complexes, e.g. [16, 17, 18, 3]
and the literature of L-systems, all reviewed from the present
operator algebra point of view in [1].

Here we will take as a working example a highly simplified 
stochastic labelled graph grammar (SLGG)
for microtubule dynamics including treadmilling, bundling/zippering,
and katanin-mediated severing, in cytoskleleton dynamics as it appears in current plant biology.
		
\subsection{MT stochastic graph grammar}

A diagrammatic presentation of an MT graph grammar,
with subscripts for the rule-local arbitrary but consistent numbering of
vertices in left- and right-hand side graphs of each rule,
is below.
Discrete parameters will include
a six-valued categorical label 
$s \in \{\operatorname{internal}, \operatorname{grow\_end}, \operatorname{retract\_end}, \operatorname{junct} \}$ 
( or $s \in \{\circ, \bullet, \blacksquare, \blacktriangle \}$ )
for status as interior segment,
growth-capable end segment, retraction-capable end segment,  
or bundling junction segment
respectively.


\begin{equation}
\begin{split}
&  \text{// Rule 1:}  \; \text{Treadmilling growth} \\
&  
\left( 
\begin{diagram}[size=1em]
\; 
\CIRCLE_1 & & &  \\
\end{diagram}
\right) 
\llangle
(\text{${ \boldsymbol x}$}_1,\text{${ \boldsymbol u}$}_1)
\rrangle  
 \longrightarrow  
\left( 
\begin{diagram}[size=1em]
\; 
\Circle_1 & \rTo(2,0) &  
\CIRCLE_2 & \\
\end{diagram}
\right) 
%
\llangle
(\text{${ \boldsymbol x}$}_1,\text{${ \boldsymbol u}$}_1),(\text{${ \boldsymbol x}$}_2,\text{${ \boldsymbol u}$}_2)
\rrangle  
\\  &  \quad \quad \text{\boldmath $\mathbf{with}$} \ \ 
	\hat{\rho }_{\text{grow}}( [\text{Y}_g])
	\mathcal{N}( \text{${ \boldsymbol x}$}_1 -  \text{${ \boldsymbol x}$}_2 ; L \text{${ \boldsymbol u}$}_1, \sigma ) 
	  \mathcal{N}_{|\text{${ \boldsymbol u}$}_2|=1}( \text{${\boldsymbol u}$}_2; \text{${ \boldsymbol u}$}_1 ), \epsilon) , \\
&  \text{// Rule 2:}  \; \text{Treadmilling retraction} \\
&  
\left( 
\begin{diagram}[size=1em]
\; 
\blacksquare_1 & \rTo(2,0) & 
\Circle_2 &  \\
\end{diagram}
\right) 
\llangle
(\text{${ \boldsymbol x}$}_1,\text{${ \boldsymbol u}$}_1),(\text{${ \boldsymbol x}$}_2,\text{${ \boldsymbol u}$}_2)
\rrangle  
  \longrightarrow  
\left( 
\begin{diagram}[size=1em]
 &  
 \blacksquare_2 & \\
\end{diagram}
\right) 
\llangle
(\text{${ \boldsymbol x}$}_2,\text{${ \boldsymbol u}$}_2)
\rrangle  
\\  &  \quad \quad \text{\boldmath $\mathbf{with}$} \ \ 
	\hat{\rho }_{\text{retract}}( [\text{Y}_r]) \\
&  \text{// Rule 3:}  \; \text{Collision-induced bundling or zippering} \\
&  
\left( 
\begin{diagram}[size=1em]
\; \Circle_1& \rTo(2,0) & & 
\Circle_2 & \rTo(2,0) & 
\Circle_3 & \\
& & &&&  \\ 
& 
\CIRCLE_4  & & \\
\end{diagram}
\right) 
\llangle
(\text{${ \boldsymbol x}$}_1,\text{${ \boldsymbol u}$}_1),(\text{${ \boldsymbol x}$}_2,\text{${ \boldsymbol u}$}_2),
  (\text{${ \boldsymbol x}$}_3,\text{${ \boldsymbol u}$}_3),(\text{${ \boldsymbol x}$}_4,\text{${ \boldsymbol u}$}_4)
\rrangle  \\
& \quad \longrightarrow   
\left( 
\begin{diagram}[size=1em]
\; \Circle_1 & \rTo(2,0) & & 
\blacktriangle_2 & \rTo(2,0) & 
\Circle_3 & \\
& & \ruTo &&&  \\ 
& 
\; \Circle_4 & & \\
\end{diagram}
\right) 
%
\llangle
(\text{${ \boldsymbol x}$}_1,\text{${ \boldsymbol u}$}_1),(\text{${ \boldsymbol x}$}_2,\text{${ \boldsymbol u}$}_2),
  (\text{${ \boldsymbol x}$}_3,\text{${ \boldsymbol u}$}_3),(\text{${ \boldsymbol x}$}_4,\text{${ \boldsymbol u}$}_4)
\rrangle  
\\  & \quad \quad \text{\boldmath $\mathbf{with}$} \ \ 
	 \hat{\rho }_{\text{bundle}} (| \text{$ {\boldsymbol u}_2$} \cdot \text{${\boldsymbol u}_4$} |/ |\cos \theta_{\text{crit}}|) 
			\exp(-| \text{${\boldsymbol x}$}_2 - \text{${\boldsymbol x}$}_4 |^2/2 L^2) \\
&  \text{// Rule 4:}  \; \text{Katanin-induced severing} \\
&  
\left( 
\begin{diagram}[size=1em]
\; \Circle_1& \rTo(2,0) &  
\Circle_2 & \rTo(2,0) & 
\Circle_3 & \\
\end{diagram}
\right) 
\llangle
(\text{${ \boldsymbol x}$}_1,\text{${ \boldsymbol u}$}_1),(\text{${ \boldsymbol x}$}_2,\text{${ \boldsymbol u}$}_2),
  (\text{${ \boldsymbol x}$}_3,\text{${ \boldsymbol u}$}_3)
\rrangle  \\
& \quad \longrightarrow   
\left( 
\begin{diagram}[size=1em]
\; \Circle_1 & \rTo(2,0) &  
\CIRCLE_2 &  & \blacksquare_4 & \rTo(2,0) & 
\Circle_3 & \\
\end{diagram}
\right) 
%
\llangle
\text{${ \boldsymbol x}$}_1,\text{${ \boldsymbol u}$}_1),(\text{${ \boldsymbol x}$}_2,\text{${ \boldsymbol u}$}_2),
  (\text{${ \boldsymbol x}$}_3,\text{${ \boldsymbol u}$}_3),(\text{${ \boldsymbol x}$}_4,\text{${ \boldsymbol u}$}_4)
\rrangle  
\\  & \quad \quad \text{\boldmath $\mathbf{with}$} \ \ 
	 \hat{\rho }_{\text{sever}}([\text{katanin}]) 
	       \mathcal{N}(  \text{${ \boldsymbol x}$}; \mathbf{0} , \sigma_{\text{broad}} ) 
		\delta_{\text{Dirac}}(| \text{${ \boldsymbol u}$}|-1 )  ) \\
\end{split}
\label{bundling_rule_2}
\end{equation}

Here $Y_g$ is a diffusible MT growth factor such as tubulin itself, or a catalyst or regulator of 
tubulin polymerization and/or nucleation, such as (perhaps) XMAP215 [19] 
and $Y_r$ plays the same role for catastrophe/retraction.

In working out the commutators we will drop the propensity functions $\rho$,
but they just multiply the results, with appropriate variable identifications.

\subsubsection{Selected MT commutator calculations}

The commutator calculations for this minimal MT graph grammar's Lie algebra can be outlined as follows.

$[\hat{W}_2,\hat{W}_1]$:

$\hat{W}_2 \cdot \hat{W}_1$:
Shared same-label vertex sets run over by $H$ and their mappings under $h$ are:
$\varnothing$; $\{(1 \mapsto 2)\}$.

$\hat{W}_1 \cdot \hat{W}_1$:
Shared same-label vertex sets run over by $H$ and their mappings under $h$ are:
$\varnothing$.

$H=\varnothing$ always cancels in the commutator.

More detailed work lets us calculate $[\hat{W}_2,\hat{W}_1]=$ by abusing notation slightly:
\begin{equation}
\begin{split}
[\hat{W}_2,\hat{W}_1]
&= 
[
\left( 
\begin{diagram}[size=1em]
\; 
\blacksquare_{1^{\prime}} & \rTo(2,0) & 
\Circle_{2^{\prime}} &  \\
\end{diagram}
\right) 
  \longrightarrow  
\left( 
\begin{diagram}[size=1em]
 &  
 \blacksquare_{2^{\prime}} & \\
\end{diagram}
\right) 
\; , \;
(
\begin{diagram}[size=1em]
\; 
\CIRCLE_1 & & &  \\
\end{diagram}
)
\longrightarrow
(
\begin{diagram}[size=1em]
\; 
\Circle_1 & \rTo(2,0) &  
\CIRCLE_2 & \\
\end{diagram}
)
]
\\ & 
\simeq \quad
\left( 
\begin{diagram}[size=1em]
\; 
\blacksquare_{1^{\prime}} & \rTo(2,0) & 
\CIRCLE_{1} &  \\
\end{diagram}
\right) 
  \longrightarrow  
\left( 
\begin{diagram}[size=1em]
\; 
\blacksquare_{1} & \rTo(2,0) & 
\CIRCLE_{2} &  \\
\end{diagram}
\right) 
\end{split}
\end{equation}
which is just a renumbering of the same graph, 
which provided that the 
model-specific rules of MT representation are respected
by the other grammar rules, should be equivalent to the identity operator.
The corresponding full $W=\hat{W}-D$ operator should therefore
(by Corollary 1 to Theorem 1) be equivalent to 
the zero operator, using this model-specific extension of the equivalence relation in 
Equation~(\ref{particle-equivalence}).

\vspace{10pt}
$[\hat{W}_3,\hat{W}_1]$:

$\hat{W}_3 \cdot \hat{W}_1$:
Shared same-label vertex sets run over by $H$ and their mappings under $h$ are:
$\varnothing$ ; 
$\{(1 \mapsto 1^{\prime})\}$ ; $\{(1 \mapsto 2^{\prime})\}$ ; $\{(1 \mapsto 3^{\prime})\}$;
$\{(1 \mapsto 1^{\prime}), (2 \mapsto 4^{\prime})\}$ ; $\{(1 \mapsto 2^{\prime}) ,(2 \mapsto 4^{\prime})\}$ ; 
$\{(1 \mapsto 3^{\prime}),  (2 \mapsto 4^{\prime})\}$.

$\hat{W}_1 \cdot \hat{W}_3$:
Shared same-label vertex sets run over by $H$ and their mappings under $h$ are:
$\varnothing$.

$H=\varnothing$ always cancels in the commutator.

\begin{equation}
\begin{split}
[\hat{W}_3,\hat{W}_1]
&= 
[
\left( 
\begin{diagram}[size=1em]
\; \Circle_{1^{\prime}}& \rTo(2,0) &  
\Circle_{2^{\prime}} & \rTo(2,0) & 
\Circle_{3^{\prime}} & \\
& & &&&  \\ 
& 
\CIRCLE_{4^{\prime}}  & & \\
\end{diagram}
\right) 
 \longrightarrow   
\left( 
\begin{diagram}[size=1em]
\; \Circle_{1^{\prime}} & \rTo(2,0) & & 
\blacktriangle_{2^{\prime}} & \rTo(2,0) & 
\Circle_{3^{\prime}} & \\
& & \ruTo &&&  \\ 
& 
\; \Circle_{4^{\prime}} & & \\
\end{diagram}
\right) 
\; , \;
(
\begin{diagram}[size=1em]
\; 
\CIRCLE_1 & & &  \\
\end{diagram}
)
\longrightarrow
(
\begin{diagram}[size=1em]
\; 
\Circle_1 & \rTo(2,0) &  
\CIRCLE_2 & \\
\end{diagram}
)
]
\\ & 
\simeq \quad
\left( 
\begin{diagram}[size=1em]
\; \Circle_{1^{\prime}}& \rTo(2,0) & & 
\Circle_{2^{\prime}} & \rTo(2,0) & 
\CIRCLE_{1} & \\
& & &&&  \\ 
& 
\CIRCLE_{4^{\prime}}  & & \\
\end{diagram}
\right) 
 \longrightarrow   
\left( 
\begin{diagram}[size=1em]
\; \Circle_{1^{\prime}} & \rTo(2,0) & & 
\blacktriangle_{2^{\prime}} & \rTo(2,0) & 
\Circle_{1} & \rTo(2,0) & &
\CIRCLE_{2} & \\
& & \ruTo &&&  \\ 
& 
\; \Circle_{4^{\prime}} & & \\
\end{diagram}
\right) 
\quad \text{(rare coincidence)}
\\ & \quad + 
\left( 
\begin{diagram}[size=1em]
\; \Circle_{1^{\prime}}& \rTo(2,0) & & 
\Circle_{2^{\prime}} & \rTo(2,0) & 
\Circle_{3^{\prime}} & \\
& & &&&  \\ 
& 
\CIRCLE_{1^{\prime}}  & & \\
\end{diagram}
\right) 
 \longrightarrow   
\left( 
\begin{diagram}[size=1em]
\; \Circle_{1^{\prime}} & \rTo(2,0) & & 
\blacktriangle_{2^{\prime}} & \rTo(2,0) & 
\Circle_{3^{\prime}} & \\
& & \ruTo &&&  \\ 
& 
\; \Circle_{2} & & \\
\ruTo(1,2) &&&  \\ 
\; \Circle_{1} & & \\
\end{diagram}
\right) 
\quad \text{(likely)} \\
\\ & \quad + 
\left( 
\begin{diagram}[size=1em]
\; \Circle_{1^{\prime}}& \rTo(2,0) & & 
\Circle_{2^{\prime}} & \rTo(2,0) & 
\CIRCLE_{1} & \\
\end{diagram}
\right) 
 \longrightarrow   
\left( 
\begin{diagram}[size=1em]
\; \Circle_{1^{\prime}} & \rTo(2,0) & & 
\blacktriangle_{2^{\prime}} & \rTo(2,0) & 
\Circle_{1} & \\
& & & \uTo & \ldTo &&  \\ 
& & & 
\Circle_{4^{\prime}} & & \\
\end{diagram}
\right) 
\quad \text{(high bending energy)} 
\\
\\ & \quad + \text{(3 terms whose LHS rely on MT syntax violations - omitted)} 
\end{split}
\end{equation}

\vspace{10pt}
$[\hat{W}_4,\hat{W}_1]$:

$\hat{W}_4 \cdot \hat{W}_1$:
Shared same-label vertex sets run over by $H$ and their mappings under $h$ are:
$\varnothing$ ; 
$\{(1 \mapsto 1^{\prime})\}$ ; $\{(1 \mapsto 2^{\prime})\}$ ; $\{(1 \mapsto 3^{\prime})\}$;

$\hat{W}_1 \cdot \hat{W}_4$:
Shared same-label vertex sets run over by $H$ and their mappings under $h$ are:
$\varnothing$;
$\{(2^{\prime} \mapsto 1)\}$ .

$H=\varnothing$ always cancels in the commutator.


\begin{equation}
\begin{split}
[\hat{W}_4,\hat{W}_1]
&= 
[
\left( 
\begin{diagram}[size=1em]
\; \Circle_{1^{\prime}}& \rTo(2,0) &  
\Circle_{2^{\prime}} & \rTo(2,0) & 
\Circle_{3^{\prime}} & \\
\end{diagram}
\right) 
 \longrightarrow   
\left( 
\begin{diagram}[size=1em]
\; \Circle_{1^{\prime}} & \rTo(2,0) &  
\CIRCLE_{2^{\prime}} &  & \blacksquare_{4^{\prime}} & \rTo(2,0) & 
\Circle_{3^{\prime}} & \\
\end{diagram}
\right) 
\; , \;
(
\begin{diagram}[size=1em]
\; 
\CIRCLE_1 & & &  \\
\end{diagram}
)
\longrightarrow
(
\begin{diagram}[size=1em]
\; 
\Circle_1 & \rTo(2,0) &  
\CIRCLE_2 & \\
\end{diagram}
)
]
\\ & 
\simeq \quad
\left( 
\begin{diagram}[size=1em]
\; \Circle_{1^{\prime}}& \rTo(2,0) &  
\Circle_{2^{\prime}} & \rTo(2,0) & 
\CIRCLE_{1} & \\
\end{diagram}
\right) 
 \longrightarrow   
\left( 
\begin{diagram}[size=1em]
\; \Circle_{1^{\prime}} & \rTo(2,0) &  
\CIRCLE_{2^{\prime}} &  & \blacksquare_{4^{\prime}} & \rTo(2,0) & 
\Circle_{1} & \rTo(2,0)& 
\CIRCLE_{2} & \\
\end{diagram}
\right) 
%
%
\\ & \quad -
\left( 
\begin{diagram}[size=1em]
\; \Circle_{1^{\prime}}& \rTo(2,0) &  
\Circle_{2^{\prime}} & \rTo(2,0) & 
\Circle_{3^{\prime}} & \\
\end{diagram}
\right) 
 \longrightarrow   
\left( 
\begin{diagram}[size=1em]
\; \Circle_{1^{\prime}} & \rTo(2,0) & 
\Circle_{2^{\prime}} & \rTo(2,0) &  
\CIRCLE_{2} & 
& \blacksquare_{4^{\prime}} & \rTo(2,0) & 
\Circle_{3^{\prime}} & \\
\end{diagram}
\right) 
\\ & \quad + \text{(2 terms whose LHS rely on MT syntax violations - omitted)} 
\end{split}
\end{equation}

The foregoing commutators can also be calculated directly by operator algebra,
bypassing the general theorems, with the same results.

\vspace{10pt}
$[\hat{W}_4,\hat{W}_3]$:
\begin{equation}
\begin{split}
[\hat{W}_4,\hat{W}_3]
&= 
[
\left( 
\begin{diagram}[size=1em]
\; \Circle_{1^{\prime}}& \rTo(2,0) &  
\Circle_{2^{\prime}} & \rTo(2,0) & 
\Circle_{3^{\prime}} & \\
\end{diagram}
\right) 
 \longrightarrow   
\left( 
\begin{diagram}[size=1em]
\; \Circle_{1^{\prime}} & \rTo(2,0) &  
\CIRCLE_{2^{\prime}} &  & \blacksquare_{4^{\prime}} & \rTo(2,0) & 
\Circle_{3^{\prime}} & \\
\end{diagram}
\right) 
\; , \;
\\ & \quad \quad
\left( 
\begin{diagram}[size=1em]
\; \Circle_{1}& \rTo(2,0) & & 
\Circle_{2} & \rTo(2,0) & 
\Circle_{3} & \\
& & &&&  \\ 
& 
\CIRCLE_{4}  & & \\
\end{diagram}
\right) 
 \longrightarrow   
\left( 
\begin{diagram}[size=1em]
\; \Circle_{1} & \rTo(2,0) & & 
\blacktriangle_{2} & \rTo(2,0) & 
\Circle_{3} & \\
& & \ruTo &&&  \\ 
& 
\; \Circle_{4} & & \\
\end{diagram}
\right) 
] \simeq \ldots
\end{split}
\end{equation}

$\hat{W}_3 \cdot \hat{W}_4$:
Shared same-label vertex sets run over by $H$ and their mappings under $h$ are:
$\varnothing$;
$\{(1^{\prime} \mapsto 1)\}$ ; $\{(1^{\prime} \mapsto 2)\}$ ; $\{(1^{\prime} \mapsto 3)\}$;
$\{(3^{\prime} \mapsto 1)\}$ ; $\{(3^{\prime} \mapsto 2)\}$ ; $\{(3^{\prime} \mapsto 3)\}$;
$\{(1^{\prime} \mapsto i), (3^{\prime} \mapsto j)\}$ where unordered sets $\{i,j\}$ are chosen without replacement from $\{1,2,3\}$ (6 possibilities);
$\{(4^{\prime} \mapsto 4)\}$ simultaneously with any of the foregoing 12 possibilities;
Thus, $4 \times 6=24$ terms.

$\hat{W}_4 \cdot \hat{W}_3$:
Shared same-label vertex sets run over by $H$ and their mappings under $h$ are:
$\varnothing$;
$\{(i \mapsto j)\}$ where $i$ is chosen from $\{1,3,4\}$ and $j$ is chosen from $\{1^{\prime},2^{\prime},3^{\prime}\}$ ($3 \times 3=9$ possibilities);
also $\{(i \mapsto k), (j \mapsto l)\}$ where 
unordered pairs $\{i,j\}$ are chosen without replacement from  $\{1,3,4\}$ (3 possibilities)
and ordered pairs $(k,l)$ are chosen without replacement from $\{1^{\prime},2^{\prime},3^{\prime} \}$ (6 possibilities,
for a total of $3 \times 6=18$ possibilities);
$\{(1 \mapsto i), (3 \mapsto j), (4 \mapsto k) \}$ where ordered sets $\{i,j,k\}$ are chosen without replacement from $\{1^{\prime},2^{\prime},3^{\prime}\}$ 
(6 possibilities).
Thus, there are $9+18+6=33$ terms.

$H=\varnothing$ always cancels in the commutator.
Other cancellations are possible, since the scalar propensity functions multiply commutatively,
leaving at most $24+33=57$ terms.
As before, many of these terms will have no effect within a grammar that preserves inductively 
valid MT representation structures.

\subsection{Related kinds of rewrite rules}

We have analyzed the semantics of, and given examples of, stochastic labelled graph grammar (SLGG) models.
In [5] we demonstrated how to use integer-valued Object ID (OID) parameters to encode
graph grammars within stochastic parameterized grammars (SPG) comprising
parameter-bearing stochastic rewrite rules with operator algebra semantics.
Since the reverse inclusion is trivial, SLGGs and SPGs are different syntax for the same semantics;
SLGGs may be easier to write since the OID encoding step is not needed.
But [5] also showed how to add to SPGs
rules with ordinary and/or stochastic differential equation syntax 
and differential operator semantics, obtaining ``dynamical grammars'' (DGs),
a matter discussed further in [1].
DGs can be taken to be a continuum limit (in label space and in time) of SPGs.
If we allow differential equation rules together with stochastic labeled graph grammar rules
we arrive at dynamical graph grammars (DGGs), again equivalent to but easier to write than DGs. 
Many other notational conveniences are possible,
while maintaining or generalizing the operator algebra semantics.

\subsubsection{Cell complex rewrite rules}

In [1] the operator algebra semantics for a labelled-graph rewrite rule is generalized in several ways.
One of these generalizations is to cell complexes (each of some maximal dimension $d$),
which have been applied to developmental modeling [16,18].
Reference [1] also provides a constructive implementation mapping from the generalized rewrite rules
back to graph grammar rewrite rules. In principle then, the graph grammar operator algebra of our
Theorems 1 and 2 apply to these generalized settings - but whether the sum of graph grammar operators
resulting from a higher-level product is also a sum of higher-level rewrite rules, or not,
remains to be worked out.

Here we point out a useful special case for cell complex dynamics:
that if a graph can be locally embedded in $d$ dimensions
(i.e. in $d$ dimensional manifolds with ${\mathbb R}^d$ as the usual case),
in such a way that it becomes a Voronoi diagram or a power diagram (weighted Voronoi diagram),
then its label set can be augmented by the resulting node positions,
and more importantly there is a dual $d$-dimensional cell complex consisting of the 
boundaries at equal distance (in the Voronoi case) from two or more graph node positions,
together with the $d$-dimensional single-node cells they bound.
Then, local graph grammar rewrite rules will generically result in local updates
to the embedding and to the dual cell complex, inducing
local cell complex changes describable as rewrite rules.

\section{Conclusions}

We have computed the product and commutator for 
any two stochastic labelled-graph rewrite rule operators,
in a stochastic graph grammar possessing operator algebra semantics,
in structural (graph-expressed, combinatorial) form.
In this form, the product of the state-changing portions
(off-diagonal in the number basis) of two graph rewrite rule operators
is a sum, with nonnegative integer coefficients, of other such operators.
Non-negative real-valued rate multipliers are also carried along in the expected way.
The product of the full graph rewrite rule operators, 
and the commutator of off-diagonal or full rule operators,
are likewise expressed as a sum with integer-valued weights of 
other full graph rewrite rule  operators.
The results are expressed in Theorem 1 and its Corollaries,
for the case of semantics in which hanging edges are left behind,
and Theorem 2 and its Corollaries, for the case in which they aren't.

There is also a computer-implementable
constructive mapping from the resulting graph rewrite rule algebra to
many elementary two-state creation/annihilation operators.
Because the algebra is expressed in the present work
entirely in terms of operators
for graph rewrite rule operators, 
rather than in terms of the underlying
elementary two-state creation/annihilation operators,
Theorems 1 and 2 are a substantial
improvement 
in utility and perspicuity
over the corresponding Propositions 1 and 2 of [1].
As a clarifying test case,
the resulting graph-grammar level algebra was applied to an elementary
example inspired by the dynamics of cortical microtubules in plant cells,
one of a large number of structure-changing dynamical systems in
biophysics and other sciences that could be amenable to 
modeling by stochastic labelled-graph grammars.

\subsection*{Acknowledgements}

\small
The author wishes to acknowledge the hospitality of the Sainsbury Laboratory Cambridge University,
the Center for Nonlinear Studies of the Los Alamos National Laboratory,
and the Kavli Institute for Theoretical Physics at the University of California Santa Barbara;
also funding from the Leverhulme Trust, 
National Institute of Aging grant R56AG059602, and Human Frontiers Science Program grant  HFSP - RGP0023/2018.
\par 
\normalsize

\section*{References}

\hangindent=3em \noindent 
[1]
Eric Mjolsness,
``Prospects for Declarative Mathematical Modeling of Complex Biological Systems''. 
Bulletin of Mathematical Biology,  Vol. 81, Issue 8, pp 3385-3420, August 2019.

\hangindent=3em \noindent 
[2]
Michael L. Blinov, James R. Faeder, Byron Goldstein and William S. Hlavacek.
BioNetGen: software for rule-based modeling of signal transduction based on the interactions of molecular domains.
Bioinformatics,
Vol. 20 no. 17, pages 3289-3291, 2004.

\hangindent=3em \noindent 
[3]
Eric Mjolsness, David H. Sharp, and John Reinitz, 
``A Connectionist Model of Development''.
Journal of Theoretical Biology, vol 152 no 4, pp. 429-454, 1991. 

\hangindent=3em \noindent 
[4]
P. Prusinkiewicz, M. S. Hammel, and E. Mjolsness, 
 ``Animation of Plant Development''.
SIGGRAPH '93 Conference Proceedings, ACM 1993.

\hangindent=3em \noindent 
[5]
Eric Mjolsness and Guy Yosiphon,
``Stochastic Process Semantics for Dynamical Grammars'', 
Annals of Mathematics and Artificial Intelligence, 47(3-4) August 2006.

\hangindent=3em \noindent 
[6]
M. Doi, Journal of Physics A: Mathematical and General 9, 1465
(1976).

\hangindent=3em \noindent 
[7]
M. Doi, Journal of Physics A: Mathematical and General 9, 1479
(1976).

\hangindent=3em \noindent 
[8]
D. C. Mattis and M. L. Glasser, Rev. Mod. Phys. 70, 979 (1998).

\hangindent=3em \noindent 
[9]
Eric Mjolsness,
``Towards Measurable Types for Dynamical Process Modeling Languages''. 
Proceedings of the 26th Conference on Mathematical Foundations of Programming Semantics (MFPS 2010).
Electronic Notes in Theoretical Computer Science (ENTCS), Elsevier, vol. 265, pp. 123-144, 6 Sept. 2010. 

\hangindent=3em \noindent 
[10]
Eric Mjolsness, 
``Time-ordered product expansions for computational stochastic systems biology''. Physical Biology, v 10, 035009, June 2013.

\hangindent=3em \noindent 
[11]
Richard P. Feyman,
``Quantum mechanical computers''
Foundations of Physics,
Volume 16, Issue 6, pp 507-531. June 1986.
See Hamiltonian on p. 517.

\hangindent=3em \noindent 
[12]
Eric Mjolsness, 
``Symbolic Neural Networks Derived from Stochastic Grammar Domain Models'', 
in Connectionist Symbolic Integration, eds. R. Sun and F. Alexandre, Lawrence Erlbaum Associates, 1997. 

\hangindent=3em \noindent 
[13]
Nicolas Behr, Vincent Danos, Ilias Garnier
``Combinatorial Conversion and Moment Bisimulation for Stochastic Rewriting Systems''
arXiv:1904.07313  April 15, 2019.

\hangindent=3em \noindent 
[14]
H. Ehrig, K. Ehrig, U. Prange, and G. Taentzer. 
Fundamentals of Algebraic Graph Transformation. 
Springer-Verlag Berlin Heidelberg, 2006. 

\hangindent=3em \noindent 
[15]
Nicolas Behr, Vincent Danos, Ilias Garnier,
``Stochastic mechanics of graph rewriting''. 
Proceedings of the 31st Annual ACM/IEEE Symposium on Logic in Computer Science, 
New York City, United States. pp.46 - 55, 2016.

\hangindent=3em \noindent 
[16]
Antoine Spicher and Olivier Michel,
``Declarative modeling of a neurulation-like process''.
Bio Systems, vol 87 2-3, pp. 281-8, 2007.

\hangindent=3em \noindent 
[17]
Jean-Louis Giavitto, Antoine Spicher
``Topological rewriting and the geometrization of programming''
Physica D 237 (2008) 1302-1314.

\hangindent=3em \noindent 
[18]
Brendan Lane,
``Cell Complexes: The Structure of Space and the Mathematics of Modularity''',
PhD thesis, University of Calgary, September 2015.

\hangindent=3em \noindent 
[19]
Olivier Hamant, Daisuke Inoue, David Bouchez, Jacques Dumais, and Eric Mjolsness,
``Are microtubules tension sensors?''
Nature Communications, v10 article no. 2360, 29 May 2019.

\section*{Appendix}

\subsection*{Pure chemical reactions: Operator algebra}

If none of the graphs involved has any edges, then each rule transforms
a collection of nodes, partitioned into indistinguishable subsets by their labels,
into another such set -- and this is equivalent to a pure stochastic chemical reaction network.
The algebra of elementary creation/annihilation operators
 is the Heisenberg algebra $[a, \hat{a}]= I$ for each chemical species $i$.
 What is the algebra of the reaction rules?
 Each reaction rule or channel has off-diagonal operator
  [D 1976a, D1976b, MG 1998]:
 
 \begin{equation}
 \hat{W}_r 
	= \hat{W}_{ \{ m_i^{(r_1)} \} \rightarrow \{n_i^{(r_1)} \} } 
	 = k^{(r)} \prod_i (\hat{a}_i)^{n_i^{(r)}} (a_i)^{m_i^{(r)}} 
 \end{equation}
so a product of such operators is
\begin{equation}
\hat{W}_{r_2}  \hat{W}_{r_1} = 
 	k^{(r_2)}  k^{(r_1)} \prod_i (\hat{a}_i)^{n_i^{(r_2)}} (a_i)^{m_i^{(r_2)}} 
	 (\hat{a}_i)^{n_i^{(r_1)}} (a_i)^{m_i^{(r_1)}} 
 \end{equation}
		
The middle two terms
$(a_i)^{m_i^{(r_2)}}  (\hat{a}_i)^{n_i^{(r_1)}}$
can be put into canonical form by mapping the Heisenberg algebra
into generating functions, $a \rightarrow \partial_x, \hat{a} \rightarrow x \times \ldots$:
\begin{equation}
\begin{split}
a^m \hat{a}^n \rightarrow& [ (\partial_x)^m x^n] \circ f(x) =  (\partial_x)^m (x^n f(x) ) 
\\ &
= \sum_{l=0}^{\min(m,n)} \binom{m}{l}
	({\partial_x}^l x^n  ) 
	({\partial_x}^{m-l}  f(x) ) 
\\ &
= \sum_{l=0}^{\min(m,n)} \binom{m}{l}
	(n)_l  x^{n-l}
	({\partial_x}^{m-l}  f(x) ) 
\\ &
\leftarrow 
	 \sum_{l=0}^{\min(m,n)} \frac{(m)_l (n)_l}{l!}
	 	\hat{a}^{n-l}  a^{m-l}
\end{split}
\end{equation}
where $(n)_l \equiv n!/(n-l)!$ for $l \leq n$. If we define also $n_l \equiv 0$ for $l > n$ then we can increase or remove the upper limit,
e.g. replace $\min$ by $\max$

Then
\begin{equation}
\begin{split}
\hat{W}_{r_2}  \hat{W}_{r_1} &= 
 	k^{(r_2)}  k^{(r_1)} 
	\prod_i \Big[
		\sum_{l_i=0}^{\min(m_i^{(r_2)}, n_i^{(r_1)})} 
		\frac{(m_i^{(r_2)})_{l_i} (n_i^{(r_1)})_{l_i}}{l_i !}
	 	(\hat{a}_i)^{n_i^{(r_1)} + n_i^{(r_2)} -l_i}  (a_i)^{m_i^{(r_1)} + m_i^{(r_2)} -l_i} 
\Big]
\\ &=
 	k^{(r_2)}  k^{(r_1)} 
	\sum_{\{l_i=0 \ldots \min(m_i^{(r_2)}, n_i^{(r_1)}) \} } 
		\Big( \prod_i 
		\frac{(m_i^{(r_2)})_{l_i} (n_i^{(r_1)})_{l_i}}{l_i !}
		\Big) \Big[ \prod_i 
	 	(\hat{a}_i)^{n_i^{(r_1)} + n_i^{(r_2)} -l_i}  (a_i)^{m_i^{(r_1)} + m_i^{(r_2)} -l_i} 
	\Big]
\end{split}
\end{equation}
i.e.
\begin{equation}
\boxed{
\begin{split}
\hat{W}_{ \{ m_i^{(r_2)} \} \rightarrow \{n_i^{(r_2)} \} } 
\hat{W}_{ \{ m_i^{(r_1)} \} \rightarrow \{n_i^{(r_1)} \} } 
 &= 
 	k^{(r_2)}  k^{(r_1)} 
	\sum_{\{l_i=0 \ldots \min(m_i^{(r_2)}, n_i^{(r_1)}) \} } 
		\Big( \prod_i 
		\frac{(m_i^{(r_2)})_{l_i} (n_i^{(r_1)})_{l_i}}{l_i !}
		\Big)
		\\ & \quad \quad \times 
		\hat{W}_{ \{ ( m_i^{(r_1)} + m_i^{(r_2)} -l_i  ) \} \rightarrow \{ ( n_i^{(r_1)} + n_i^{(r_2)} -l_i ) \} } 
\end{split}
}
\end{equation}

Likewise 
\begin{equation}
\boxed{
\begin{split}
[\hat{W}_{ \{ m_i^{(r_2)} \} \rightarrow \{n_i^{(r_2)} \} } ,
&
\hat{W}_{ \{ m_i^{(r_1)} \} \rightarrow \{n_i^{(r_1)} \} } ] \\
 &= 
 	k^{(r_2)}  k^{(r_1)} 
	\sum_{\{l_i=0 \ldots \min(m_i^{(r_2)}, n_i^{(r_1)}) \} \wedge {\mathbf l} \neq \mathbf{0} } 
		\Bigg[
			\Big( \prod_i 
				\frac{(m_i^{(r_2)})_{l_i} (n_i^{(r_1)})_{l_i}}{l_i !}
			\Big)
			-
			\Big( \prod_i 
				\frac{(m_i^{(r_1)})_{l_i} (n_i^{(r_2)})_{l_i}}{l_i !}
			\Big)
		\Bigg]
		\\ & \quad \quad \quad \times 
		\hat{W}_{ \{ ( m_i^{(r_1)} + m_i^{(r_2)} -l_i  ) \} \rightarrow \{ ( n_i^{(r_1)} + n_i^{(r_2)} -l_i ) \} } 
\end{split}
}
\end{equation}
where ${\mathbf l} \neq \mathbf{0}$ is the particle analog of Corollaries 4 or 9 regarding the cancellation
of $H = \varnothing$ from a graph grammar commutator.

\end{document}